\documentclass[journal]{IEEEtran}

\ifCLASSINFOpdf

\else

\fi
\usepackage{geometry}
\usepackage{setspace}
\usepackage{graphicx}
\usepackage{times}
\usepackage{enumitem}
\usepackage{url}
\usepackage{amsmath}
\usepackage{amssymb}
\usepackage{multirow}

\begin{document}
\setlength{\parskip}{6pt}
\setlength{\parindent}{0pt}

\title{Bridging Brain Signals and Language: \\ A Deep Learning Approach to EEG-to-Text Decoding}

\author{Mostafa~El Gedawy,~\IEEEmembership {Omnia~Nabil,}
	Omar~Mamdouh,~\IEEEmembership {Mahmoud~Nady,}
	Nour~Alhuda Adel, Ahmed Fares \\ ~\IEEEmembership{Department of Computer Science and Engineering}
	\\~\IEEEmembership{Egypt Japan University Of Science and Technology, Borg Al Arab, Egypt
	}
}

\maketitle

\begin{abstract}
	Brain activity translation into human language delivers the capability to revolutionize machine-human interaction while providing communication support to people with speech disability. Electronic decoding reaches a certain level of achievement yet current EEG-to-text decoding methods fail to reach open vocabularies and depth of meaning and individual brain-specific variables. We introduce a special framework which changes conventional closed-vocabulary EEG-to-text decoding approaches by integrating subject-specific learning models with natural language processing methods to resolve detection obstacles. This method applies a deep representation learning approach to extract important EEG features which allow training of neural networks to create elaborate sentences that extend beyond original data content. The ZuCo dataset analysis demonstrates that research findings achieve higher BLEU, ROUGE and BERTScore performance when compared to current methods. The research proves how this framework functions as an effective approach to generate meaningful and correct texts while understanding individual brain variations. The proposed research aims to create a connection between open-vocabulary Text generation systems and human brain signal interpretation for developing efficacious brain-to-text systems. The research produces interdisciplinary effects through innovative assistive technology development and personalized communication systems which extend possibilities for human-computer interaction in various settings.
\end{abstract}


\IEEEpeerreviewmaketitle

\begin{center}
\end{center} \begin{figure*}[t]
\centering
\includegraphics[width=0.8\textwidth]{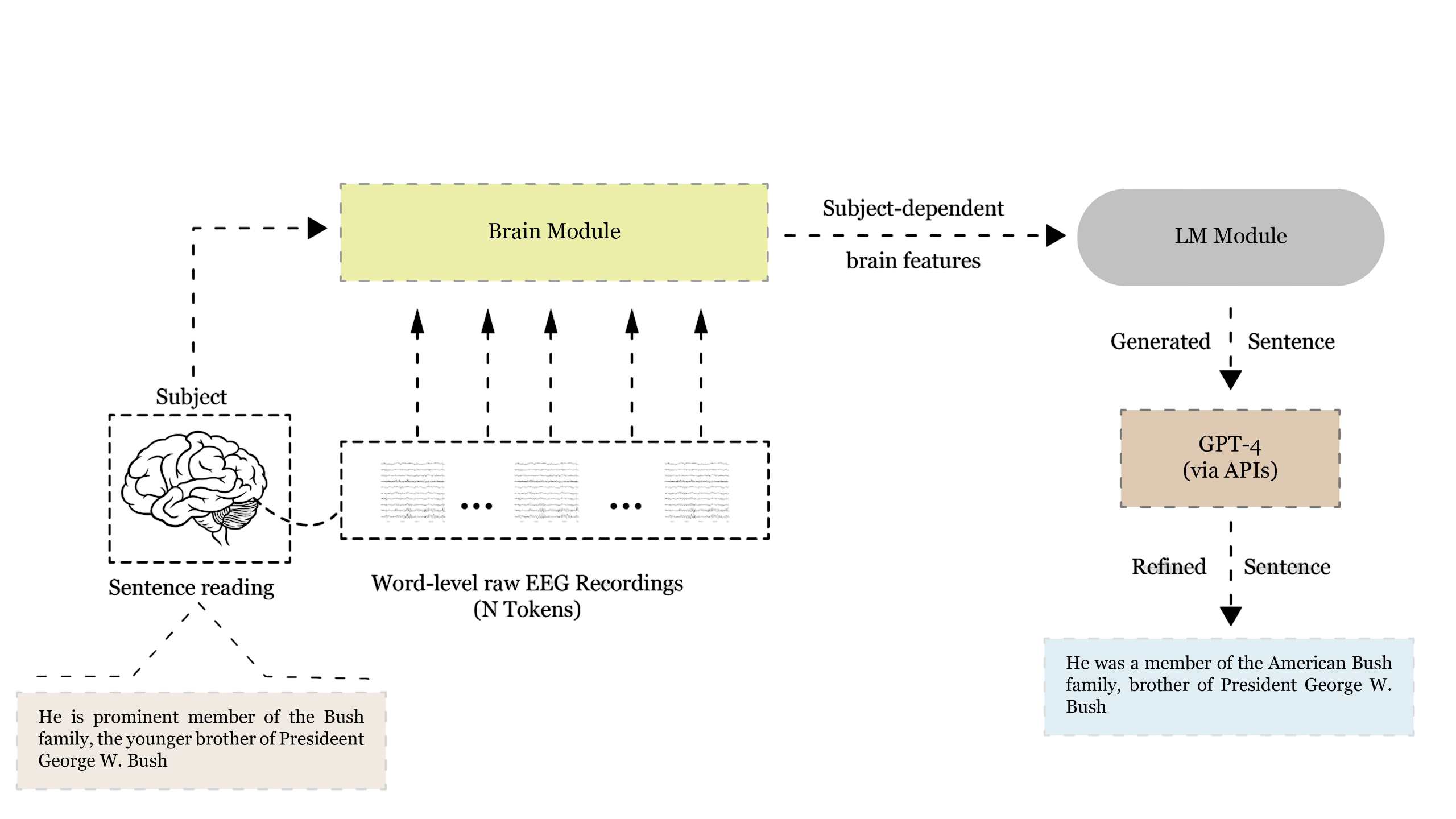} 
\vspace{-10pt} 
\caption\normalsize\centering{  
	\textbf{Input:} The subject reads sentences while EEG signals are recorded.
	\\  \textbf{Brain Module:} The EEG signals corresponding to individual words are processed to extract subject-dependent features.
	\\ \textbf{Language Module:} The features are input to a BART-based language model, which generates a preliminary sentence.
	\\\textbf{Refinement Module:} A GPT-4-based model refines the preliminary sentence to improve its fluency and accuracy.
	\\\textbf{Output:} The final sentence is generated, aligned with the target ground truth. The figure illustrates an example where the true sentence is "He is a member of the Bush family, the younger brother of President George W. Bush." he final decoded sentence accurately aligns with this target, showcasing the system’s effectiveness.}
\label{fig:workflow}
\end{figure*}

\section{Introduction}

\IEEEPARstart{T}{he} brain of human beings performs two remarkable neuropsychological activities by creating language while reshaping abstract ideas into organized speech. The inability to produce speech becomes a severe handicap for people suffering from amyotrophic lateral sclerosis (ALS) and locked-in syndrome along with those who have experienced traumatic brain injuries. The inability to produce speech known as muteness develops from three main areas: biological factors, psychological factors, and neurological factors through stroke-induced brain damage along with Parkinson’s disease or autism or Down syndrome \cite{unknown-author-2022A, lyberg-ahlander-2018}. According to both sources, the language impairment known as aphasia affects 1 in 272 Americans following strokes because stroke damage to language-processing brain areas leads to this condition in 25-40

Patients with aphasia face difficulties expressing basic needs because of which they experience frustration together with isolation along with severe depression. The severe effects of speech impairments become apparent through the case of U.S. Congresswoman Gabby Giffords who suffered aphasia following a gunshot wound that damaged her brain \cite{unknown-author-2022B}. Assistive technology allowed Dr. Stephen Hawking with ALS to communicate thus demonstrating the necessity of innovative speech restoration solutions as he used similar tools \cite{servick-2021}. Current brain-computer interface technologies demonstrate they can translate neural signals into spoken speech which brings new possibilities to patients with paralysis and anarthria (inability to use speech muscles) \cite{servick-2021}. A research team has restored the ability for paralyzed patients to produce sentences using brain signal analysis which provides a communication rate of 18 words per minute \cite{servick-2021}.

The technological advancement of brain-computer interfaces enables assistance with natural language skills through neural signal decoding to produce speech or text forms \cite{Moses2021, Willett2023}. Electroencephalography (EEG) demonstrates excellence among non-invasive imaging solutions with its cost-effective features and mobility and extremely quick time-based core while simultaneously monitoring brain processes in real-time \cite{Hollenstein2018, Panachakel2021}. Scientists face tremendous hurdles in their attempt to develop a system that transforms EEG signals into meaningful language since technical limitations exist alongside neural data complexity \cite{Duan2023, Feng2023}.

\subsection{Context and Background of the Research Problem}
The human brain's ability to generate language and transform abstract thought into structured speech is one of the most complex processes in neuroscience. For individuals with severe speech or motor impairments such as those caused by amyotrophic lateral sclerosis (ALS), locked-in syndrome, or traumatic brain injuries, this natural ability is altered, leaving them isolated from verbal communication. Brain-computer interfaces (BCI) have emerged as a groundbreaking solution to assist innate language skills by decoding neural activity into text or speech \cite{Moses2021, Willett2023}.

Among non-invasive neuroimaging techniques, electroencephalography (EEG) stands out due to its affordability, portability, and high temporal resolution, which allows real-time monitoring of brain activity \cite{Hollenstein2018, Panachakel2021}. However, translating EEG signals into coherent language remains a formidable challenge, constrained by technical limitations and the inherent complexity of interpreting neural data \cite{Duan2023, Feng2023}.

Existing EEG-to-text systems predominantly operate within \textbf{closed-vocabulary paradigms}, where decoding is restricted to a predefined set of words or phrases \cite{Wang2022, Pandarinath2017}. These systems achieve moderate success in controlled laboratory settings but do not address real-world communication needs, which demand \textbf{open-vocabulary flexibility}. In addition, EEG signals are notoriously noisy, with low spatial resolution and significant variability between individuals \cite{Jeng2020}. Traditional approaches often overlook this \textbf{inter-subject variability}, leading to models that perform poorly when applied to new users.

Many recent breakthroughs in natural language processing (NLP) have revolutionized text generation, namely the rise of \textbf{pre-trained language models} (e.g., BERT, BART, GPT) \cite{Devlin2018, Lewis2019}. These models show their worth at mapping sequences to meaning, but their integration with EEG data remains nascent \cite{Caucheteux2022}. The challenge being aligning low-dimensional, noisy EEG signals with the high-dimensional embeddings of language models \cite{Tang2023}.

To democratize access to this technology, our research extends beyond algorithmic innovation to include \textbf{a deployable web platform}. This platform operationalizes our EEG-to-text framework, allowing users to upload EEG data, generate text through our model, and convert the output into natural-sounding speech \cite{OpenAI2023}. A detailed description of the web platform design along with accessibility features appears in Appendix A.

\subsection{Problem Statement}
Despite progress in EEG-based BCIs, critical gaps hinder their practical adoption:
\begin{enumerate}[label=(\arabic*)]
	\item \textbf{Closed-Vocabulary Constraints}: Restrictive word sets fail to accommodate dynamic, open-ended dialogue \cite{Wang2022}.
	\item \textbf{Subject-Specific Variability}: EEG signal heterogeneity across individuals is rarely addressed, leading to poor model generalization \cite{Jeng2020}.
	\item \textbf{Semantic Deficits}: Existing methods prioritize syntactic accuracy but neglect the meaningfulness of decoded text \cite{Gauthier2018}.
	\item \textbf{Integration Challenges}: Integrating EEG data in modern NLP models is complicated due to the high-frequency and highly noisy nature of EEG signals  \cite{Caucheteux2022}.
	\item \textbf{Accessibility Barriers}: Most EEG-backed NLP systems are designed for research labs, lacking the user-friendliness necessary for real-world deployment \cite{Nieto2022}.
\end{enumerate}

These limitations accentuate the need for a paradigm shift in EEG-to-text decoding—one that embraces open-vocabulary generation, personalization, semantic fidelity, and practical accessibility.

\subsection{Research Objectives}
This thesis aims to advance EEG-to-text decoding through a dual focus on algorithmic innovation and real-world implementation. The specific objectives are:
\begin{enumerate}[label=(\arabic*)]
	\item \textbf{Develop a Modular End-to-End Framework}:
	      \begin{itemize}
	      	\item \textbf{Brain Module}: Design a subject-specific feature extraction pipeline using bi-directional GRUs and transformer encoders \cite{Vaswani2017}.
	      	\item \textbf{Language Module}: Integrate BART to generate open-vocabulary text from EEG-derived features \cite{Lewis2019}.
	      	\item \textbf{Refinement Module}: Employ GPT-4 for post-hoc semantic and grammatical correction \cite{OpenAI2023}.
	      \end{itemize}
	\item \textbf{Build a User-Centric Web Platform}:
	      \begin{itemize}
	      	\item Develop an intuitive interface for uploading EEG data and receiving text-to-speech outputs \cite{Nieto2022}.
	      	\item Ensure compatibility with diverse EEG datasets and hardware \cite{Hollenstein2018}.
	      \end{itemize}
	\item \textbf{Enable Voice Synthesis}:
	      \begin{itemize}
	      	\item Convert decoded text into natural speech using state-of-the-art voice synthesis models (e.g., WaveNet, Tacotron) \cite{Moses2021}.
	      \end{itemize}
	\item \textbf{Validate System Efficacy}:
	      \begin{itemize}
	      	\item Evaluate performance using syntactic and semantic metrics (BLEU, ROUGE, BERTScore) \cite{Lin2004, Zhang2019}.
	      	\item Conduct user studies with individuals with communication impairments \cite{Willett2023}.
	      \end{itemize}
	\item \textbf{Address Ethical and Practical Challenges}:
	      \begin{itemize}
	      	\item Implement data privacy measures to protect neural data \cite{Pandarinath2017}.
	      	\item Optimize computational efficiency for real-time processing \cite{Duan2023}.
	      \end{itemize}
\end{enumerate}

\subsection{Significance of the Study}
This work bridges theoretical innovation with tangible societal impacts:
\begin{itemize}
	\item \textbf{Assistive Communication}: Enables individuals with speech impairments to express thoughts via brain activity \cite{Moses2021}.
	\item \textbf{Real-World Accessibility}: The web platform democratizes BCI technology for broader clinical and home use \cite{Nieto2022}.
	\item \textbf{Multimodal Interaction}: Integrates voice synthesis for natural conversation and accessibility \cite{Willett2023}.
	\item \textbf{Advancements in Neuroscience and NLP}: Offers insights into neural encoding of language and cross-modal learning \cite{Caucheteux2022}.
	\item \textbf{Scalability and Adaptability}: Supports multilingual customization and regional dialect adaptation \cite{Wang2022}.
\end{itemize}

\subsection{Methodology and Web Platform Workflow}
\subsubsection{EEG Data Acquisition and Preprocessing}
\begin{itemize}
	\item \textbf{Input}: Users upload preprocessed EEG signals in pickle file format \cite{Hollenstein2018}.
	\item \textbf{Preprocessing}: Noise removal, normalization, and segmentation into word-level intervals \cite{Duan2023}.
\end{itemize}

\subsubsection{Brain Module: Subject-Specific Feature Extraction}
\begin{itemize}
	\item Bi-Directional GRU processes temporal EEG data \cite{Cho2014}.
	\item Subject-Specific Layer adjusts weights based on user neural patterns \cite{Jeng2020}.
	\item Brain Transformer Encoder extracts global dependencies \cite{Vaswani2017}.
\end{itemize}

\subsubsection{Language Module: Open-Vocabulary Text Generation}
\begin{itemize}
	\item BART fine-tuning generates preliminary text predictions \cite{Lewis2019}.
	\item GPT-4 refines outputs for semantic and grammatical accuracy \cite{OpenAI2023}.
\end{itemize}

\subsubsection{Voice Synthesis and Delivery}
\begin{itemize}
	\item Text-to-Speech (TTS) converts final text into human-like speech \cite{Moses2021}.
\end{itemize}

\section{Related Work}
Brain-to-speech and brain-to-text decoding related work can be identified by three main entities that they are capturing: namely; motor imagery based, overt speech based, and inner speech based. Several different BCI devices are available and exist respectively, and these can be broadly defined as the usage of Electroencephalography (EEG), Electrocorticography (ECoG), and functional Magnetic Resonance Imaging (fMRI) \cite{Panachakel2021}.

Motor imagery-based systems, such as for instance, point-and-click \cite{Jarosiewicz2015} and imaginary handwriting \cite{Willett2021}, have high accuracy but very slow typing rate. The simultaneous complementary event-related potential (ERP)-based P300 speller, steady state visually evoked potential (SSVEP), as well as code-modulated visual evoked potential (c-VEP) paradigms have been developed that employ neural signals in translating brain signals to text \cite{Lee2018}. The ERP system operates by monitoring neural activity in the brain that occurs in response to sensory events, while SSVEP and c-VEP paradigms exploit correlated neuronal potentials by means of visual evoked potentials, utilizing their different levels of alienation from the user in the course of information circulation, transmission of instant or in-/out-deck via operators or indicators and susceptibility to fatigue.

Decoding or synthesis of speech over speech activity which is real speech activity, this is considered as the overt speech-based method. The method is characterized by faster communication rates \cite{Makin2020} as compared to the existing modes. This technique necessitates the participation of subjects in vocal exercises during the neuro recording process \cite{Anumanchipalli2019} or the subjects have to perform the mental work of saying the sentence aloud \cite{Brigham2010}.

While developing such options, this makes the system itself become language-dependent. There are significant differences in pronunciations in various languages. Inner speech-based methods attempt to resolve the language articulation dependency by decoding language from the speculative speech and reading of text \cite{Defossez2023, Nieto2022}.

A significant limitation of the majority of the approaches under discussion is the restriction of using small closed vocabularies, with a low and limited number of unique words \cite{Brigham2010, Pandarinath2017}. In conclusion, the majority of the current communication through language approaches involve invasive devices such as (ECoG) \cite{Willett2021} or less accessible non-invasive devices like fMRI \cite{Nieto2022}.

This situation makes it more difficult to collect large datasets of the speech and put in place of methods to help paralyzed people who cannot speak anymore. However, the most recent research endeavors are trying to decode inner speech by opening the vocabulary and also using non-invasive technology \cite{Defossez2023, Nieto2022}.

Launching the possibility of similarly decoding studies of brain-to-text conversion of the inner voice. We study EEG signal representation learning, inter-subject variability, human judgment at the sentence level of generated sentences \cite{Wang2022}.

\section{Methodology}

The goal of open vocabulary EEG-to-Text decoding is to interpret human brain activity—captured using EEG recordings—and translate it into coherent, meaningful text. This involves decoding high-dimensional data from brain signals recorded in real-time while a person is reading English sentences. The task leverages advanced machine learning models capable of understanding the complexity of brain signals and their relationship to linguistic constructs.

\subsection{Dataset}

Zurich Cognitive Language Processing Corpus( ZuCo) is the dataset employed. It combines eye shadowing and electroencephalography( EEG) data from healthy persons who speaks English as their first language( forming from Canada, USA, UK or Australia, South Africa) while they read rulings in the language. It contributes to exploration on mortal reading and language appreciation by analysing brain exertion and eye movement.

Table~\ref{tab:compare_zuco1_zuco2}, ~\ref{tab:Details_zuco_1}, ~\ref{tab:Details_zuco_2} illustrate the differences between the two dataset versions. There are 140 positive, 137 negative, and 123 neutral sentences in the 400 Normal reading (Sentiment) selected sentences at ZuCo 1.0. The 739 sentences that were chosen from the Wikipedia corpus at ZuCo 2.0, it was chosen because it provides annotations of semantic relations. Relation identification is a sophisticated semantic problem that calls for intricate mental operations. Seven of the relation types that were initially specified were included: 
\begin{itemize}
	\item political affiliation
	\item education
	\item founder
	\item wife/husband
	\item job title
	\item nationality
	\item employer
\end{itemize}

\begin{table*}[!t]
	\renewcommand{\arraystretch}{1.3}
	\caption{\large Comparison between ZuCo 1.0 and ZuCo 2.0}
	\label{tab:compare_zuco1_zuco2}
	\centering
	\fontsize{12pt}{12pt}\selectfont
	\begin{tabular}{|l|l|l|}
		
		\hline
		ZuCo 1.0                              & ZuCo 2.0                                     \\ \hline
		12 persons                            & 18 persons                                   \\ \hline
		5 females and 7 males                 & 10 females and 8 males                       \\ \hline
		1107 sentences:                       & 739 sentences from Wikipedia:                \\
		400 Normal reading (Sentiment)        & 349 normal reading paradigm                  \\
		300 Normal reading (Wikipedia)        & 390 sentences task specific reading paradigm \\
		407 Task-specific reading (Wikipedia) &                                              \\ \hline
		21,629 words                          & 15,138 words                                 \\ \hline
		Ages between 22 and 54 years          & Ages between 25 and 42 years                 \\ \hline
		Three reading tasks                   & Two reading tasks                            \\ \hline
		Two conventional reading tasks        & A normal reading task                        \\ \hline
		One task-specific reading exercise    & A specific reading task                      \\ \hline
	\end{tabular}
\end{table*}

\begin{table*}[!t]
	\renewcommand{\arraystretch}{1.3}
	\caption{\large  Details about ZuCo 1.0: sentence and word information}
	\label{tab:Details_zuco_1}
	\centering
	\fontsize{12pt}{12pt}\selectfont
	\begin{tabular}{|l|l|l|l|}
		\hline
		P.O.C                              & NR1                 & NR2                 & TSR                    \\ \hline
		sentences                          & 400                 & 300                 & 407                    \\ \hline
		sentence length (mean (SD), range) & 17.7 (8.29), 3–43 & 21.2 (10.5), 5–62 & 20.06  (10.09), 5–62 \\ \hline
		total words                        & 7079                & 6386                & 8164                   \\ \hline
		word types                         & 3080                & 2657                & 2995s                  \\ \hline
		word length (mean (SD), range)     & 6.97 (2.71), 1-26   & 6.7 (2.6), 1-29     & 6.69 (2.58), 1-21      \\ \hline
	\end{tabular}
\end{table*}

\begin{table*}[!t]
	\renewcommand{\arraystretch}{1.3}
	\caption{\large  Details about ZuCo 2.0: sentence and word information}
	\label{tab:Details_zuco_2}
	\centering
	\fontsize{12pt}{12pt}\selectfont
	\begin{tabular}{|l|l|l|}
		\hline
		P.O.C                              & NR               & TSR              \\ \hline
		sentences                          & 349              & 390              \\ \hline
		sentence length (mean (SD), range) & 19.6 (8.8), 5-53 & 21.3 (9.5), 5-53 \\ \hline
		total words                        & 6828             & 8310             \\ \hline
		word types                         & 2412             & 2437             \\ \hline
		word length (mean (SD), range)     & 4.9 (2.7), 1-29  & 4.9 (2.7), 1-21  \\ \hline
	\end{tabular}
\end{table*}

The reading paradigm task required participants to identify whether or not a certain relation type was present in the sentence. The ultimate objective is to substitute physiological activity data captured from human sentence reading for this assignment. Text understanding and annotation can be decoded using this data, which includes eye-tracking and brain activity signals. Natural language processing and machine learning are the main uses for the dataset. 
By processing eye-position data, the Eye Link 1000 tracker can detect blinks and fixations. Individual fixation gaze position (x,y) items are included in the dataset. Regarding the monitor coordinates, the coordinates were given in pixels; the upper left corner of the screen was (0,0), and the down/right was positive.The eye-tracking characteristics shown below have been extracted:

\begin{itemize}
	\item Gaze duration (GD): The total number of fixations on the current word during the first pass reading (excluding regressions and rereading) before the eye leaves the word. It begins when a participant's gaze lands on a word for the first time and concludes when they shift to another word. Among the fixations are: 
	      \begin{itemize}
	      	\item The first fixation on the word.
	      	\item Saccade to another part of the word means the eyes move to focus on another part of the word.
	      	\item Refixation means the eyes stop to process more details of the word.
	      	\item Final Saccade to the next word means the eyes move to the next word.
	      \end{itemize}
	\item Total reading time (TRT): the sum of all fixation durations on the current word, including regressions.
	\item First fixation duration (FFD): the duration of the first fixation on a word.
	\item Single fixation duration (SFD): the duration of the first and only fixation on the current word.
	\item Go-past time (GPT): the sum of all fixations prior to progressing to the right of the current word, including regressions to previous words that originated from the current word. For each of these eye-tracking features we have additionally computed the pupil size. Furthermore, we have extracted the number of fixations and mean pupil size for each word and sentence.
\end{itemize}

Important definitions for eye tracking data:
\begin{itemize}
	\item Fixations: when the eyes stop and focus on a word, and they are separated by saccades. Longer fixations suggest more effort is needed to process information.
	\item Saccades: are rapid eye movements that occur between fixations.
	\item Fixations and saccades make up the process of reading.
	      
\end{itemize}

For the EEG signals the total channels are 128 (Geodesic Hydrocel system) , the used for analysis are 105 EEG channels were used from the scalp recordings, 9 EOG channels (for eye movement artifact removal) and 14 channels lying mainly on the neck and face were discarded before data analysis. 

\begin{figure}[!h]
	\renewcommand{\arraystretch}{1.3}
	
	\centering
	\caption\normalsize\centering{Shows the scalp and eeg channels, the recording reference was at Cz}
	\label{fig:scalp_channels}
	\includegraphics[width=0.5\textwidth, height=0.26\textheight]{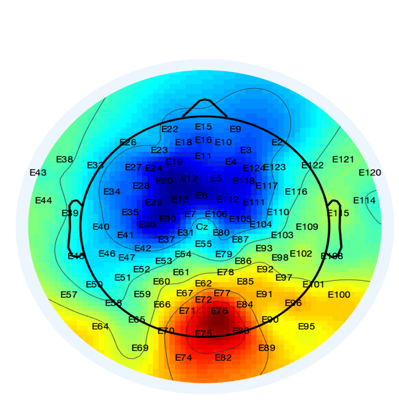}     \vspace{-10pt} 
	
\end{figure}

Figure~\ref{fig:scalp_channels} shows recording reference was at Cz. it is located at the central electrode location on the scalp. This central placement minimizes signal biases and ensures balanced referencing across scalp regions. Other electrodes measure signals relative to Cz, an electrode near your forehead might measure +50 µV another electrode at the back of your head might measure -30 µV these values are relative to Cz, which is 0. Referring to Cz makes it easier to detect meaningful patterns in brain activity during reading because all signals have a common baseline.

The analysis focused on oscillatory power across frequency bands, with time-series data also shared. This was achieved by applying band-pass filtering to continuous EEG signals during the task. There are five independent frequency bands:

\begin{itemize}
	\item Theta bands: Help understand attention and memory processes during reading.
	      \begin{itemize}
	      	\item theta1 (4–6 Hz)
	      	\item theta2 (6.5–8 Hz)
	      \end{itemize}
	\item Alpha bands: Show levels of engagement with the text.
	      \begin{itemize}
	      	\item alpha1 (8.5–10 Hz)
	      	\item alpha2 (10.5–13 Hz)
	      \end{itemize}
	\item Beta bands: Active during intense mental activity.
	      \begin{itemize}
	      	\item beta1 (13.5–18 Hz) 
	      	\item beta2 (18.5–30 Hz)
	      \end{itemize}
	\item Gamma bands: Demonstrate how different aspects of language processing are unified into coherent understanding.
	      \begin{itemize}
	      	\item gamma1 (30.5–40Hz) 
	      	\item gamma2 (40–49.5Hz)
	      \end{itemize}
\end{itemize}

The sampling rate of EEG signals and Eye-tracking records is 500Hz this makes EEG signals and eye-tracking systems record data in synchronized way this ensure that the timing of eye fixations aligns with the corresponding EEG signals. Uses gaze duration to determine the relevant time window that the eye is fixating on a specific word to segmenting the EEG data based on eye fixation periods. This helps to understand the neural processes related to reading specific words as when the eye-tracking data show the word that reader was focusing on, The EEG tells us the brain activity at that exact moment. 

The sampling rate is 500Hz as maximum frequency rate of 8 frequency band is 100Hz of gamma band, Band pass of datasets is 0.1 to 100Hz. Nyquist theorem is used to calculate the minimum sampling rate. The minimum required sampling rate $f_{\text{sampling}}$ must be:

\begin{equation}
	f_{\text{sampling}} \geq 2 \times f_{\text{max}}
\end{equation}

\begin{equation}
	f_{\text{sampling}} \geq 2 \times 100 = 200 \, \text{Hz}
\end{equation}

\noindent where \( f_{\text{max}} \) is the highest frequency of interest.

Theoretically, EEG signals up to 100 Hz can be recorded at a sampling rate of 200 Hz. To improve accuracy, accommodate for any noise, and enable more robust preprocessing, a higher sample rate (500 Hz) is employed.

\textbf{The datasets used for the project:}

\begin{figure}[!h]
	\renewcommand{\arraystretch}{1.3}
	\centering
	\caption\normalsize\centering{Shows how the content of each dataset in pickel file looks like}
	\label{fig:dataset_content}
	\includegraphics[width=0.5\textwidth, height=0.9\textheight]{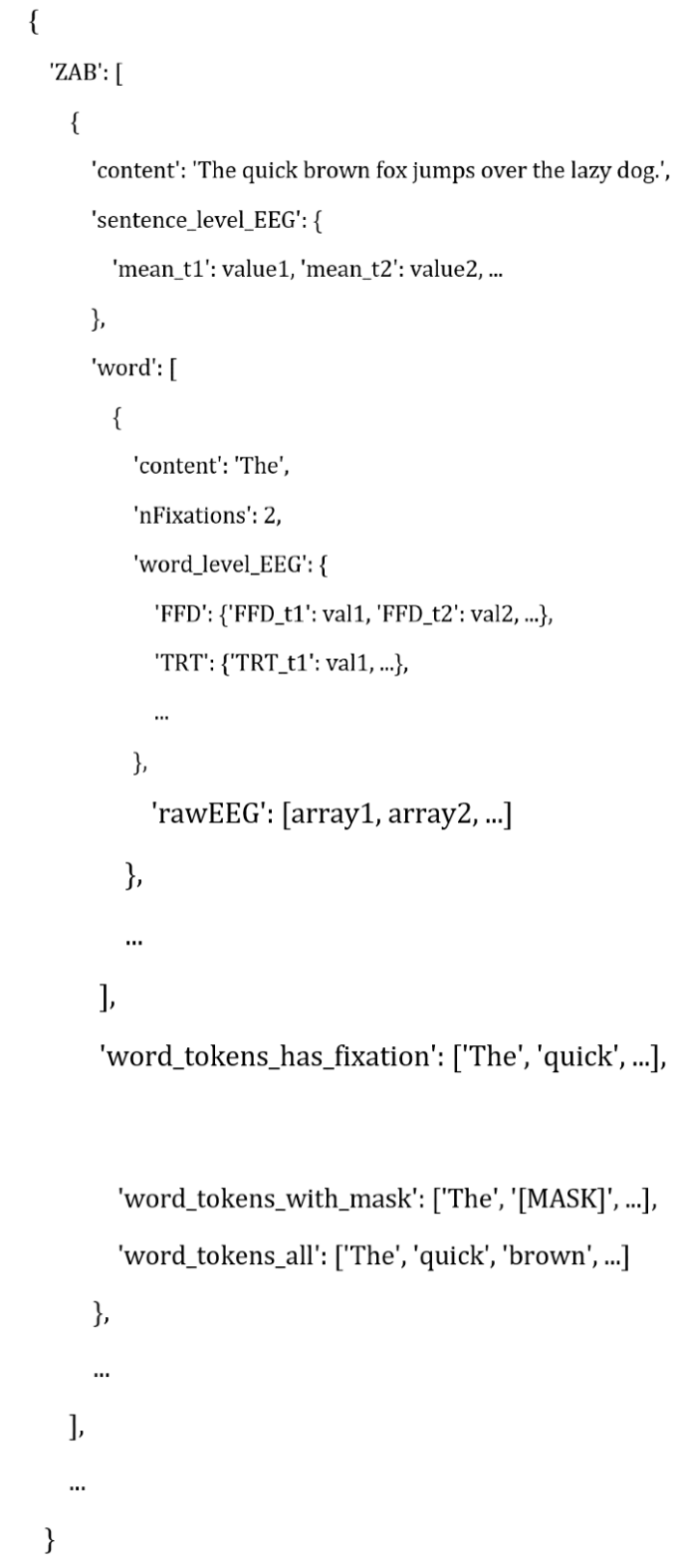}     \vspace{-10pt} 
\end{figure}

Figure~\ref{fig:dataset_content} shows the content of each dataset file (its extension is pickle). Each subject represents the key in the big dictionary ‘{….}’ (dataset) and the value is list ‘[….]’ of dictionaries, each dictionary in the list contains a sentence and its EEG signals and EEG signals during (gaze duration (GD), total reading time (TRT), first fixation duration (FFD) ). Data was collected from a total of 30 subjects - 12 subjects in ZuCo v1.0 and 18 subjects in ZuCo v2.0 in natural reading tasks. The 3 tasks used: 

\begin{itemize}
	\item Normal Reading (Sentiment)
	\item Normal Reading (Wikipedia) v1.0
	\item Normal Reading (Wikipedia) v2.0
\end{itemize}

The shape of dataset of task 1 is (12,400,7):
\begin{itemize}
	\item 12: number of subjects.
	\item 400: number of sentences for each subject.
	\item 7: data of each sentence ['content', 'sentence level EEG', 'answer EEG','word', 'word tokens has fixation', 'word tokens with mask', 'word tokens all']
\end{itemize}
The shape of dataset of task 2 v1.0 is (12,300,7) \\
The shape of dataset of task 2 v2.0 is (18, 349,6):
\begin{itemize}
	\item 6: data of each sentence ['content', 'sentence level EEG', 'word', 'word token has fixation', 'word tokens with mask', 'word tokens all']
\end{itemize}

\subsection{EEG-to-Text Decoding}

The EEG signals are represented mathematically as a two-dimensional matrix:
\begin{equation}
	X \in \mathbb{R}^{C \times T}
\end{equation}

Here:
- \(C\): Number of EEG channels (electrodes used in recording brain activity).
- \(T\): Number of time steps over which the data is recorded. Each time step captures neural activity at a particular moment.

Each EEG sequence, \(X\), corresponds to a particular subject, \(s\), and belongs to a dataset of distinct subjects, denoted by \(S\). The goal of decoding is to predict a text sentence \(Y\) based on the brain signals \(X\). Each sentence \(Y\) consists of individual tokens \(y_n\), where \(y_n\) represents an English word drawn from an open vocabulary \(V\) that contains all possible words in the language.

Sequence-to-Sequence Modeling

This decoding process is framed as a Sequence-to-Sequence problem, where:
- The input sequence is the EEG signal \(X\).
- The output sequence is the corresponding text sentence \(Y\).

The relationship between \(X\) and \(Y\) is defined by a decoding function \(f\):
\begin{equation}
	f : \{C \times T\} \times S \rightarrow V
\end{equation}

The function \(f\) takes two inputs: the EEG data \(X\) and the subject identifier \(s\). It outputs a predicted text sentence \(Y\). This mapping enables the translation of neural activity into coherent, readable text.

The sentence \(Y\) is expressed as:

\begin{equation} Y = f(X, s)\end{equation}

This means that the function \(f\) generates a sentence \(Y\) based on the recorded EEG signals \(X\) and the specific subject \(s\). During training, the model learns to optimize this function to improve prediction accuracy.

Probabilistic Decoding

To achieve accurate decoding, the system maximizes the probability of the text sentence \(Y\) given the EEG signals \(X\):

\begin{equation} p(Y \mid X) = \prod_{n=1}^N p(y_n \in V \mid X, y_{<n})\end{equation}

Here:
- \(N\): Number of tokens in the sentence \(Y\).
- \(y_n\): The \(n\)-th token in the sentence.
- \(y_{<n}\): All tokens generated before \(y_n\).

The expression highlights the sequential nature of the decoding process. Each token \(y_n\) is generated by considering:
1. The EEG signals \(X\), which encode neural activity.
2. The tokens \(y_{<n}\), which provide the context from previously generated words.

By iterating through all tokens in the sentence, the model constructs the complete text \(Y\). This probabilistic formulation ensures that the generated sentence is coherent and aligns with the brain activity recorded in \(X\).

Importance of Open Vocabulary

The use of an open vocabulary \(V\) is a crucial aspect of this task. Unlike classification tasks with a fixed set of predefined outputs, an open vocabulary allows the model to generate any word in the English language. This flexibility is essential for producing meaningful text that reflects the subject's reading material or thought process.

For example, if a subject is reading a sentence containing a rare or specialized word, the model can include that word in its prediction. This adaptability makes the system suitable for a wide range of applications, from communication aids for individuals with disabilities to brain-computer interfaces for creative expression.

Breakdown of the Decoding Process

1. Signal Processing: The EEG signals \(X\) are preprocessed to remove noise and extract meaningful features. These signals represent neural activity over time, captured across multiple channels.
2. Contextual Modeling: The model examines \(X\) and considers previously generated tokens \(y_{<n}\) to predict the next token \(y_n\). This ensures that the generated text maintains grammatical and semantic consistency.
3. Sequential Generation: The model iteratively predicts each token in the sentence, building \(Y\) step by step. The probability of each token \(y_n\) is conditioned on \(X\) and the context provided by \(y_{<n}\).

By combining advanced neural architectures with probabilistic modeling, this approach enables the translation of complex neural activity into human-readable text. It demonstrates the potential of EEG-based systems for natural language generation and highlights the transformative possibilities of brain-computer interfaces. 
Overview of the Proposed Architecture

The proposed architecture is a well-structured system for decoding EEG signals into coherent text, comprising two key components:
\begin{enumerate}
	\item \textbf{Brain Module:} Implements a representation learning method tailored for encoding EEG signals. This module converts the high-dimensional EEG signals into meaningful features.
	\item \textbf{Language Modeling Module:} Transforms the EEG-based features into readable text. It leverages two advanced language models:
	      \begin{itemize}
	      	\item \textbf{BART:} A pre-trained model that produces EEG-to-Text translations.
	      	\item \textbf{GPT-4:} Enhances the generated sentences by refining grammar, structure, and fluency.
	      \end{itemize}
\end{enumerate}

\noindent \textbf{Key Points of the Training Process:}
\begin{itemize}
	\item Training is divided into \textbf{two stages}.
	\item Components enclosed in \textbf{dashed boxes} in the architectural diagram are trained, while those in \textbf{solid boxes} remain fixed.
\end{itemize}

\subsection{Training Process: Two Stages}

\subsubsection{Training Stage 1: Brain Module Training}
In this stage, the Brain module is trained to align raw word-level EEG signals with word tokens derived from the BART language model.

\subsubsection{Steps and Key Objectives:}
\begin{itemize}
	\item \textbf{Input:} EEG signals (\(X\)) are processed at the word level. These signals represent brain activity.
	\item \textbf{Representation Mapping:} A learnable features module processes the EEG signals, accounting for subject-specific nuances and encoding them into features (\(Z\)).
	\item \textbf{Objective:} The goal is to map the learned EEG representations (\(Z\)) to the token embeddings from the locked, pre-trained BART model (\(BART_{te}^{enc}\)).
\end{itemize}

\noindent This alignment is achieved by minimizing the \textbf{Mean Square Error (MSE) Loss}, which ensures that the Brain module learns to produce EEG features (\(Z\)) close to the target token embeddings (\(BART_{te}^{enc}\)) using MSE regression loss \(L_{MSE}(BART_{te}^{enc}, Z)\).

\begin{equation}
	\min_{f_{brain}} L_{MSE}(BART_{te}^{enc}, f_{brain}(X))
\end{equation}

\begin{itemize}
	\item \(L_{MSE}\): Measures the squared differences between the predicted EEG representation \(f_{brain}(X)\) and the BART token embeddings \(BART_{te}^{enc}\).
	\item \(f_{brain}(X)\): The Brain module function that processes EEG signals (\(X\)) to produce the encoded representation (\(Z\)).
	\item \(BART_{te}^{enc}\): Token embeddings from the locked, pre-trained BART model.
\end{itemize}
\begin{center}
\end{center}
\begin{figure*}[t]
	\centering
	\includegraphics[width=0.75\textwidth]{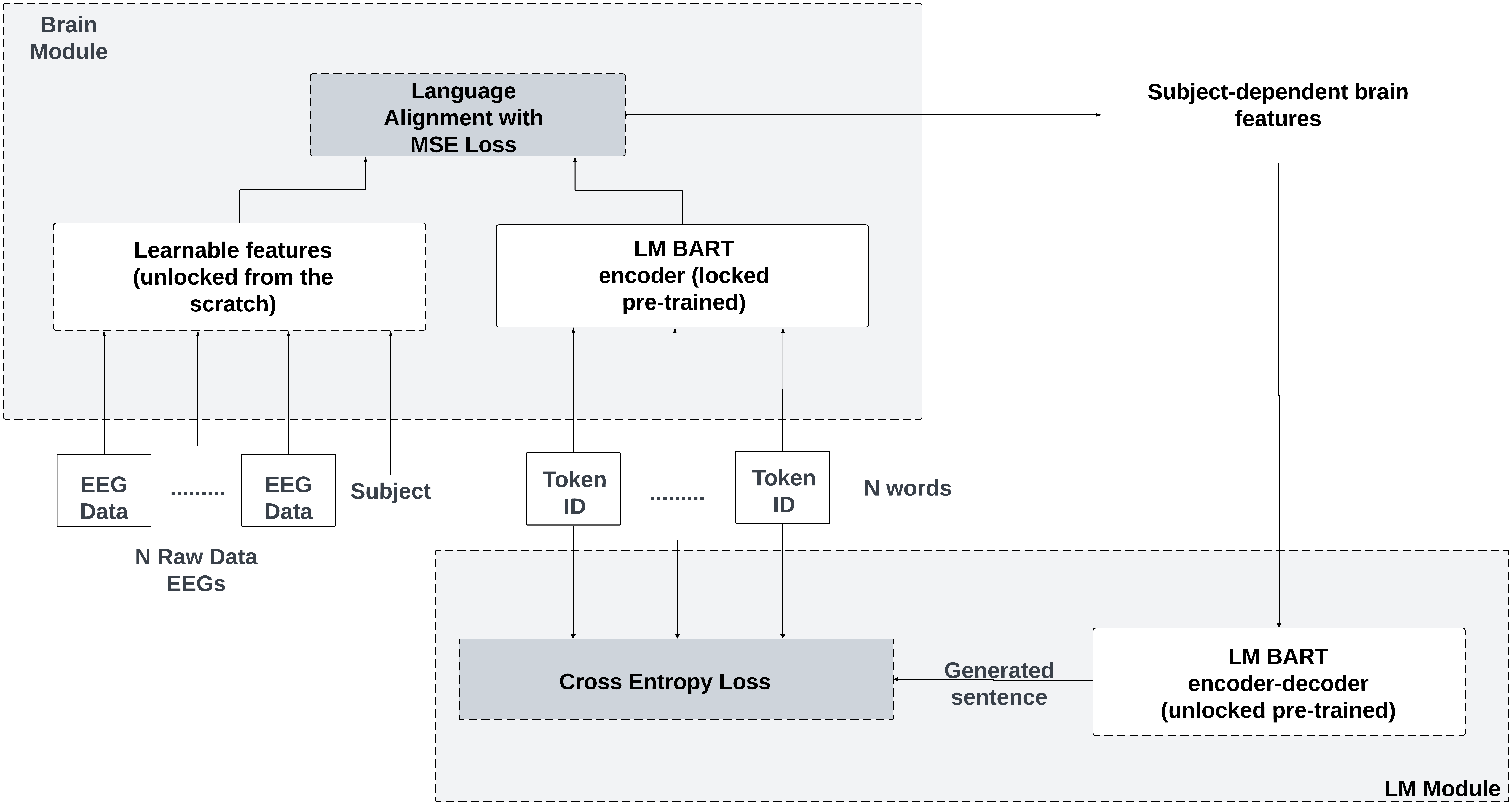} 
	\vspace{-10pt} 
	\caption\normalsize\centering{  
		\textbf Illustrates a comprehensive view of the proposed system for translating EEG signals into open vocabulary text. Initially, raw EEG signals at the word level are input into the Brain module, which is responsible for extracting deep and meaningful representations tailored for encoding the raw EEG data. Following this, the system employs a Language Modeling (LM) module to transform the processed EEG data into textual sentences. This is achieved by utilizing the pre-trained BART language model. The diagram highlights the distinction between trainable and non-trainable components: modules enclosed in dashed boxes undergo training during the system's development, while those within solid boxes remain fixed and untrained throughout the process.}
	\label{fig:workflow}
\end{figure*}
\noindent \textbf{Outcome:} After this stage, the Brain module learns to generate subject-specific EEG features that align with word-level token embeddings from the BART model.
\subsection{Training Stage 2: Language Model Fine-Tuning}
In the second stage, the pre-trained BART language model is fine-tuned to generate meaningful English sentences from the EEG signal representations (\(Z\)) produced by the Brain module.
\subsubsection{Steps and Key Objectives:}
\begin{itemize}
	\item \textbf{Input:} The EEG representations (\(Z\)) serve as the initial embeddings for the BART encoder-decoder.
	\item \textbf{Decoding Process:}
	      \begin{enumerate}
	      	\item Each EEG representation (\(Z\)) is treated as a word-level input.
	      	\item The BART encoder-decoder processes these embeddings to generate text sequences.
	      	\item The final hidden states of the BART decoder are passed through a Multi-Layer Perceptron (MLP), which predicts the English tokens (\(y_n\)) from the BART vocabulary (\(V\)).
	      \end{enumerate}
	\item \textbf{Objective:} Minimize the \textbf{Cross-Entropy Loss}, ensuring the predicted tokens (\(y_n\)) match the actual tokens in the target text sentence.
\end{itemize}
\begin{equation}
	L_{rec} = - \sum_{n=1}^N \log p(y_n \in V)
\end{equation}

\begin{itemize}
	\item \(L_{rec}\): Measures the negative log probability of the correct tokens (\(y_n\)) belonging to the vocabulary (\(V\)).
	\item \(N\): The number of tokens in the sentence.
	\item \(p(y_n \in V)\): The predicted probability of the token \(y_n\) being part of the vocabulary (\(V\)).
\end{itemize}

\noindent \textbf{Outcome:} The BART language model learns to translate the EEG signal representations into meaningful sentences. The generated sentences are later refined by GPT-4 for enhanced coherence and readability.

\subsection{Breakdown of Modules}

\subsubsection{Brain Module}
\begin{itemize}
	\item \textbf{Function:} Encodes EEG signals into meaningful representations (\(Z\)).
	\item \textbf{Training:} Trained to align its outputs with the token embeddings of a pre-trained BART model (\(BART_{te}^{enc}\)) using MSE Loss.
\end{itemize}

\subsubsection{Language Modeling Module}
\begin{itemize}
	\item \textbf{BART:} A pre-trained encoder-decoder model, fine-tuned to generate text from EEG representations.
	\item \textbf{GPT-4:} Refines the generated sentences to improve grammatical accuracy, coherence, and fluency.
\end{itemize}

\subsection{Learnable Features Module}

The \textbf{Learnable Features Module} is a key component of the Brain module, responsible for extracting subject-specific features from raw EEG signals. It processes the high-dimensional and complex EEG data to produce latent brain representations tailored to individual subjects.

\subsubsection{Input Data}
\begin{itemize}
	\item \textbf{EEG Signals:} A sequence of word-level raw EEG signals denoted as:
	      
	      \begin{equation}
	      	\\X = \{x_0, x_1, \dots, x_M\} \in \mathbb{R}^{C \times T}
	      \end{equation}
	      where:
	      \begin{itemize}
	      	\item \( C \): Number of EEG channels.
	      	\item \( T \): Duration of the signal in time steps.
	      	\item \( M \): Number of words in the sequence.
	      \end{itemize}
	\item \textbf{Subject Information:} Each sequence of EEG signals is associated with a subject \( s \in S \), where \( S \) represents the set of all subjects.
\end{itemize}

\subsubsection{Output Data}
The output is a latent subject-specific brain representation:

\begin{equation}
	Z = \{z_0, z_1, \dots, z_M\} = f_{brain}(X) \in \mathbb{R}^M
\end{equation}

Here, \( Z \) captures meaningful features that account for subject variability, improving the model's ability to decode personalized EEG data.

\subsection{Architecture and Processing Steps}

The Learnable Features Module processes the EEG data through several stages, leveraging advanced neural network techniques to extract and refine features.

\begin{figure*}[t]
	\centering
	\includegraphics[width=0.55\textwidth]{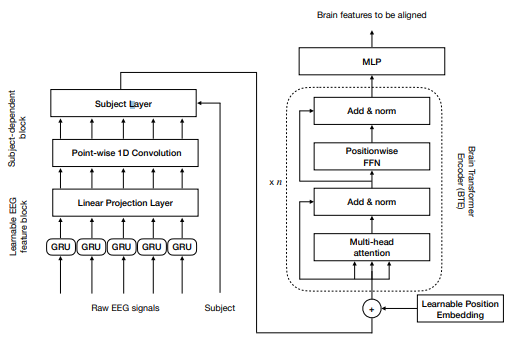} 
	\vspace{-10pt} 
	\caption\normalsize\centering{  
		The Learnable features module consists of (1) Bi-Directional Gated Recurrent Unit (GRU), (2) Linear Projection Layer, (3) 1D Pointwise Convolution
		Transformer Encoder), (4) Subject-Specific Layer ,     \\(5)  Brain Transformer Encoder (BTE) and (6) Residual Multi-Layer Perceptron (MLP) .}
	\label{fig:workflow_2}
\end{figure*}

\subsubsection{Bi-Directional Gated Recurrent Unit (GRU)}
\begin{itemize}
	\item \textbf{Purpose:} The GRU processes the multi-time series EEG signals in both forward and backward directions. This bidirectional approach dynamically handles varying lengths of word-level EEG signals while capturing temporal dependencies.
	\item \textbf{Output:} The GRU outputs forward (\( \overrightarrow{h_t} \)) and backward (\( \overleftarrow{h_t} \)) representations for each time step \( t \).
\end{itemize}

\subsubsection{Linear Projection Layer}
The forward and backward outputs from the GRU are concatenated and passed through a fully connected layer, transforming the representations into a unified feature vector for each word-level EEG signal.

\subsubsection{1D Pointwise Convolution}
\begin{itemize}
	\item A pointwise convolution with a kernel size of 1 is applied to the EEG features.
	\item \textbf{Purpose:}
	      \begin{itemize}
	      	\item Ensures dimensional uniformity by transforming the feature depth into \( D \), a consistent number of output channels.
	      	\item Prepares the features for subject-specific adjustments in the next step.
	      \end{itemize}
\end{itemize}

\subsubsection{Subject-Specific Layer}
\begin{itemize}
	\item \textbf{Purpose:} To account for inter-subject variability, the model learns a subject-specific row vector \( r_s \in \mathbb{R}^D \) for each subject \( s \in S \).
	\item This vector is applied along the channel dimension of the EEG features, enabling personalized decoding.
\end{itemize}

\subsubsection{Brain Transformer Encoder (BTE)}
The processed EEG features are passed through a multi-layer transformer encoder (\( BT_E \)) to capture global dependencies in the data.

\begin{itemize}
	\item \textbf{Transformer Details:}
	      \begin{itemize}
	      	\item \( L \): Number of layers in the transformer.
	      	\item \( H \): Number of attention heads in each layer.
	      	\item \( d_h \): Intermediate hidden dimension size.
	      \end{itemize}
	\item \textbf{Input Initialization:} The first input to the transformer (\( BT_E^{(0)} \)) is produced using a weight matrix \( W_{in} \in \mathbb{R}^{d_h \times l} \), combined with a learnable 1D position embedding \( P \).
	\item \textbf{Layer Operations:}
	      \begin{enumerate}
	      	\item \textbf{Self-Attention:} Captures relationships between EEG features within a sequence.
	      	\item \textbf{Feed-Forward Network:} Enhances feature extraction.
	      	\item \textbf{Normalization and Dropout:} Stabilizes training and prevents overfitting.
	      \end{enumerate}
\end{itemize}

\subsubsection{Residual Multi-Layer Perceptron (MLP)}
The final output of the transformer (\( BT_E^{(L)} \)) is passed through an MLP composed of two fully connected layers, producing the latent brain representations (\( z_m \)).

\subsection{Sentence Refinement During Inference}
    
\subsubsection{Purpose}
The \textbf{Sentence Refinement} step uses a pre-trained GPT-4 model to enhance text quality, improving comprehensibility, grammatical accuracy, and fluency.

\subsubsection{Process}
\begin{enumerate}
	\item \textbf{Input:} A generated sentence \( Y \) from the Language Modeling Module.
	\item \textbf{Prompt Design:} The GPT-4 model is instructed to act as a text reconstructor with the following prompt:
	\item \textbf{Processing by GPT-4:}
	      \begin{itemize}
	      	\item The generated sentence \( Y \) is corrected for grammatical errors, repetitive words, and punctuation marks.
	      \end{itemize}
	\item \textbf{Output:} A refined text sentence with improved readability and grammatical accuracy.
\end{enumerate}

\subsubsection{Significance}
\begin{itemize}
	\item \textbf{Error Reduction:} Fixes errors such as improper grammar and punctuation.
	\item \textbf{Enhanced Fluency:} Improves sentence flow for better comprehension.
	\item \textbf{Consistency:} Ensures minimal deviation from the original intent of the generated text.
\end{itemize}

\section{Results}

Our proposed architecture follows the same structure as the current state-of-the-art models by Hamza Amrani \cite{Amrani}. As shown in Table~\ref{tab:Table_of_our_results}, the evaluation metrics achieved are a BLEU-1 score of 42.34\%, BLEU-2 score of 25.26\%, BLEU-3 score of 14.91\%, BLEU-4 score of 8.89\%, a ROUGE-1-F score of 32.66\%, a ROUGE-2-F score of 9.60\%, and a BERTScore-F of 53.53\%.

We trained the architecture with additional large language models (LLMs), namely T5 and ProphetNet. The BART model produced the best results. Decoding examples of EEG-to-text generated phrases, with and without GPT-4 refinement, are presented in Table~\ref{tab:Table_of_examples} alongside ground truth comparisons. The Bold words in  Table~\ref{tab:Table_of_examples} refer to the common words between the target sentence (ground truth) and the predicted sentence by the model. The model effectively decodes named entities, particularly those present in the training set. For instance, in example (2), it successfully decodes \textit{``Beta Kappa''} and academic achievements. However, errors persist, such as in example (3), where it misinterprets \textit{``World War II''} decorations, and in example (4), where it incorrectly renders \textit{``presidential election''}.

\begin{figure*}[!t]
	\renewcommand{\arraystretch}{1.3}
	\centering
	\caption\normalsize\centering{BLEU-1,2,3,4 Score Comparison}
	\label{fig:bleu_score}
	\includegraphics[width=\textwidth, height=0.26\textheight]{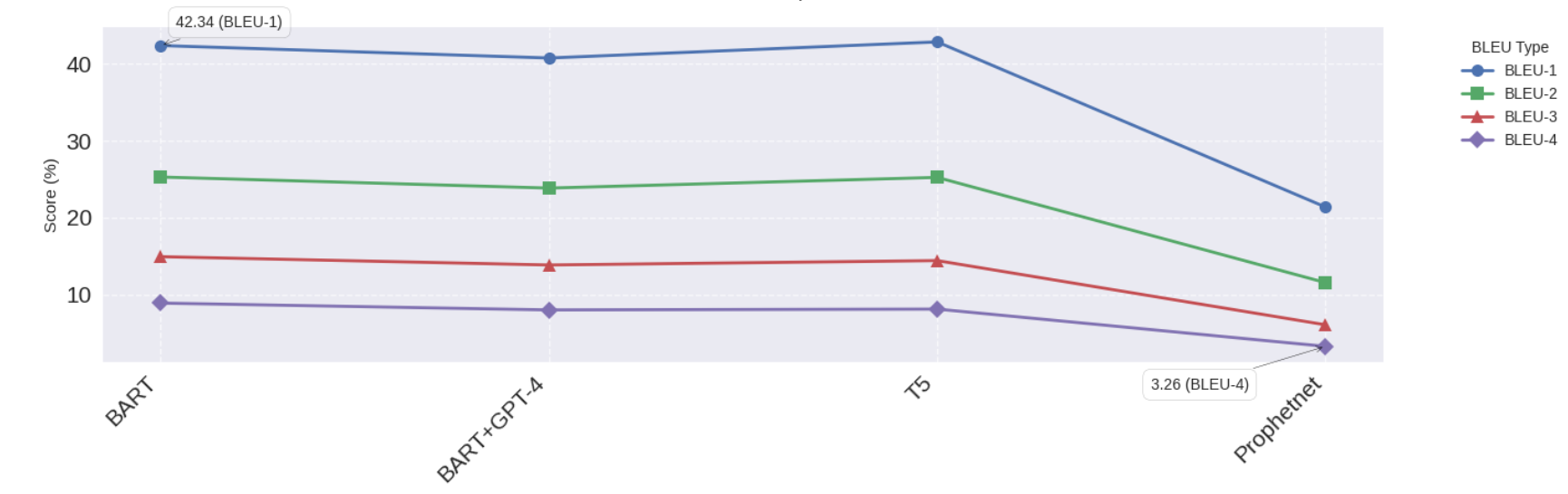} 
	\vspace{-10pt} 
	
\end{figure*}

\begin{figure*}[!t]
	\renewcommand{\arraystretch}{1.3}
	
	\centering
	\caption\normalsize\centering{ROUGE-1 Score Comparison}
	\label{fig:rouge_score}
	\includegraphics[width=\textwidth, height=0.26\textheight]{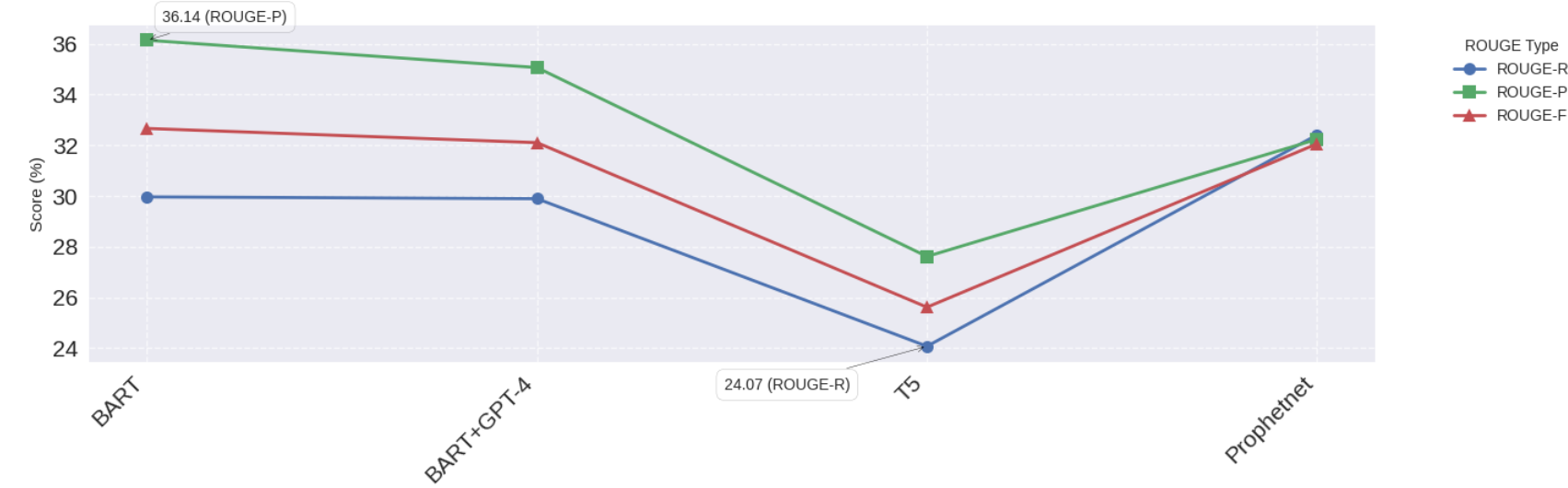}     \vspace{-10pt} 
	
\end{figure*}

\begin{figure*}[!t]
	\renewcommand{\arraystretch}{1.3}
	\centering
	\caption\normalsize\centering{ BERTScore Score Comparison}
	\label{fig:bert_score}
	\includegraphics[width=\textwidth, height=0.26\textheight]{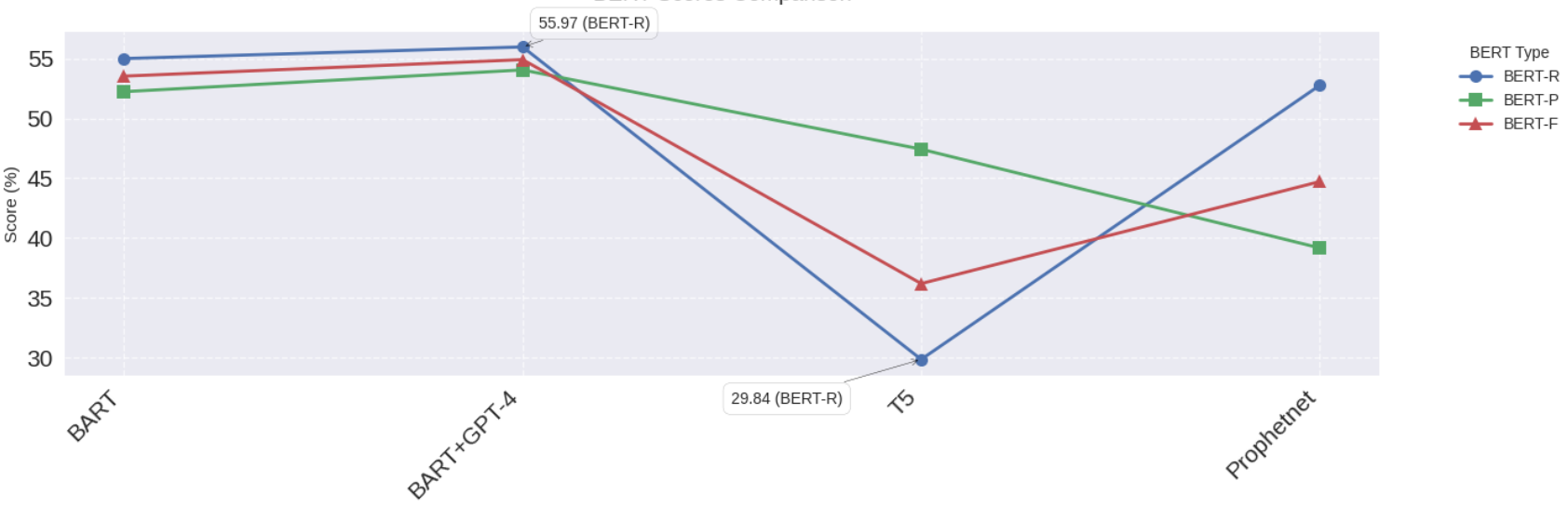} 
	\vspace{-10pt} 
	
\end{figure*}

Figure~\ref{fig:bleu_score}, ~\ref{fig:rouge_score}, ~\ref{fig:bert_score} show which model has the highest or lowest BLEU-N score, ROUGE-1 score and BERTScore. Figure ~\ref{fig:bleu_score} shows the BART model has best BLEU score and ProphetNet model has the lowest BLEU score. Figure ~\ref{fig:rouge_score} shows the BART model has best ROUGE score and T5 model has the lowst ROUGE score. Figure ~\ref{fig:bert_score} shows the BART + GPT-4 model has best BERTScore and T5 model has the lowest BERTScore. This means that BART model is almost the best.

\begin{table*}[!t]
	\renewcommand{\arraystretch}{1.3}
	\caption{\large Model evaluation on ZuCo datasets comparing BART, BART+GPT-4, T5, and ProphetNet using BLEU-N, ROUGE-1, and BERTScore metrics}
	\label{tab:Table_of_our_results}
	\centering
	\fontsize{12pt}{12pt}\selectfont
	\begin{tabular}{|l|cccc|ccc|ccc|}
		\hline
		\multirow{2}{*}{\textbf{Models}} & \multicolumn{10}{c|}{\textbf{Scores}} \\
		\cline{2-11}
		& \multicolumn{4}{c|}{\textbf{BLEU-N (\%)}} & \multicolumn{3}{c|}{\textbf{ROUGE-1 (\%)}} & \multicolumn{3}{c|}{\textbf{BERTScore (\%)}} \\
		\cline{2-11}
		           & N=1   & N=2   & N=3   & N=4  & R     & P     & F     & R     & P     & F     \\
		\hline
		BART       & 42.34 & 25.26 & 14.91 & 8.88 & 29.96 & 36.14 & 32.66 & 55.0  & 52.23 & 53.53 \\
		BART+GPT-4 & 40.72 & 23.82 & 13.83 & 7.99 & 29.89 & 35.06 & 32.1  & 55.97 & 54.05 & 54.91 \\
		T5         & 42.81 & 25.21 & 14.4  & 8.1  & 24.07 & 27.61 & 25.61 & 29.84 & 47.42 & 36.18 \\
		ProphetNet & 21.36 & 11.54 & 6.06  & 3.26 & 32.4  & 32.23 & 32.05 & 52.76 & 39.16 & 44.71 \\
		\hline
	\end{tabular}
\end{table*}

\begin{table*}[!t]
	\renewcommand{\arraystretch}{1.3}
	\caption{\large Model evaluation on ZuCo datasets in the paper of Hamza Amrani \cite{Amrani}  comparing BART, BART+GPT-4 using BLEU-N, ROUGE-1, and BERTScore metrics}
	\label{tab:Table_of_literature_results}
	\centering
	\fontsize{12pt}{12pt}\selectfont
	\begin{tabular}{|l|cccc|ccc|ccc|}
		\hline
		\multirow{2}{*}{\textbf{Models}} & \multicolumn{10}{c|}{\textbf{Scores}} \\
		\cline{2-11}
		& \multicolumn{4}{c|}{\textbf{BLEU-N (\%)}} & \multicolumn{3}{c|}{\textbf{ROUGE-1 (\%)}} & \multicolumn{3}{c|}{\textbf{BERTScore (\%)}} \\
		\cline{2-11}
		           & N=1   & N=2   & N=3   & N=4  & R     & P     & F     & R     & P     & F     \\
		\hline
		BART       & 42.75 & 25.90 & 15.66 & 9.56 & 30.60 & 36.71 & 33.28 & 55.26 & 52.62 & 53.86 \\
		BART+GPT-4 & 40.87 & 24.43 & 14.53 & 8.82 & 30.40 & 35.50 & 32.61 & 56.30 & 54.58 & 55.34 \\
		          
		\hline
	\end{tabular}
\end{table*}

\begin{table*}[!htbp]
	\renewcommand{\arraystretch}{1.3}
	\caption{\large Displays ZuCo test sentence examples with predictions made by BART, BART+GPT-4, T5, and ProphetNet models}
	\label{tab:Table_of_examples}
	\centering
	\fontsize{11pt}{12pt}\selectfont
	\begin{tabular}{|p{0.05\textwidth}|p{0.15\textwidth}|p{0.73\textwidth}|}
		\hline
		\textbf{Ex.} & \textbf{Model}  & \textbf{Text}                                                                                                                                                                                                                                                                                                                  \\
		\hline
		1            & Target sentence & He is a prominent member of the Bush family, the younger brother of President George W. Bush and the second son of former President George H. W. Bush and Barbara Bush.                                                                                                                                                        \\ \cline{2-3}
		             & BART            & was the \textbf{member member} \textbf{of} American \textbf{family}, and son brother of President George W. Bush. \textbf{the younger brother of President President George W}. W. \textbf{Bush}. former \textbf{Bush}.                                                                                                        \\ \cline{2-3}
		             & BART+GPT-4      & \textbf{He} was \textbf{a member} \textbf{of the} American \textbf{Bush family}, and \textbf{the} \textbf{younger brother of} former \textbf{President George W. Bush}.                                                                                                                                                        \\ \cline{2-3}
		             & T5              & n a figure of the administration and and brother of former Bush \textbf{W. Bush}, the father oldest of \textbf{President President George W.}W. \textbf{Bush}. his \textbf{Bush}.                                                                                                                                              \\ \cline{2-3}
		             & ProphetNet      & \textbf{he} was the former \textbf{member} \textbf{of the} american \textbf{family} and and son \textbf{brother} \textbf{of} \textbf{george george w. bush. the son son of} george \textbf{president george w w w. bush}. first \textbf{bush}.                                                                                 \\ \hline
		
		2            & Target sentence & Bush attended the University of Texas at Austin, where he graduated Phi Beta Kappa with a Bachelor's degree in Latin American Studies in 1973, taking only two and a half years to complete his work, and obtaining generally excellent grades.                                                                                \\ \cline{2-3}
		             & BART            & was \textbf{the University of} Chicago \textbf{at Austin}, \textbf{where he} was in \textbf{Beta Kappa} in \textbf{a degree} of \textbf{degree in} History \textbf{American studies}. 18. and a one classes \textbf{a half years} of \textbf{complete}. degree. and was a \textbf{excellent grades}.                           \\ \cline{2-3}
		             & BART+GPT-4      & He was at \textbf{the University} \textbf{of} Chicago and \textbf{Austin,} \textbf{where he} was in \textbf{Beta Kappa}. He completed \textbf{a degree in American} History \textbf{studies} in one \textbf{and a half years}, at 18, \textbf{and} achieved \textbf{excellent grades}.                                         \\ \cline{2-3}
		             & T5              & the ceremony of Michigan \textbf{at Austin} and \textbf{where he} was from. \textbf{Beta Kappa}. \textbf{a degree} ofs \textbf{degree in} English \textbf{American studies}. 1981. and years \textbf{a half years} \textbf{to complete}. degree. \textbf{and} he good \textbf{grades}.                                         \\ \cline{2-3}
		             & ProphetNet      & he was \textbf{the university} \textbf{of} california \textbf{at austin,} \textbf{where he} was from \textbf{beta kappa} in \textbf{a degree} of s \textbf{degree in} economics \textbf{american studies}. 1975. \textbf{and a a years a half years to} graduate \textbf{his} degree. \textbf{and} was a good \textbf{grades}. \\ \hline
		
		3            & Target sentence & Kennedy's other decorations of the Second World War include the Purple Heart, Asiatic-Pacific Campaign Medal, and the World War II Victory Medal.                                                                                                                                                                              \\ \cline{2-3}
		             & BART            & \textbf{edy} was father children include \textbf{the} year \textbf{World War} were a American \textbf{Heart}, thecentatic StarAmerican,,, \textbf{and the} American \textbf{War II Victory Medal}.                                                                                                                             \\ \cline{2-3}
		             & BART+GPT-4      & \textbf{Edy} was a father. His children \textbf{include the} American \textbf{Heart,} the Centatic Star American, \textbf{and the} American \textbf{War II Victory Medal}, which were all awarded in the year of World War II.                                                                                                 \\ \cline{2-3}
		             & T5              & s daughter \textbf{include} \textbf{the} World War. \textbf{Heart} and thetic American,,, \textbf{and the War II} medaly \textbf{Medal}.                                                                                                                                                                                       \\ \cline{2-3}
		             & ProphetNet      & he was s father great \textbf{include the} year \textbf{world war} were a stars flag, the flag american flag flag, and the bronze \textbf{war ii medal} medal.                                                                                                                                                                 \\ \hline
		
		4            & Target sentence & Following the 1980 presidential election, Bush and his family moved to Miami-Dade County, Florida.                                                                                                                                                                                                                             \\ \cline{2-3}
		             & BART            & \textbf{the} wars \textbf{election}, the was \textbf{his} wife \textbf{moved to} Florida,\textbf{Dade County, Florida}.                                                                                                                                                                                                        \\ \cline{2-3}
		             & BART+GPT-4      & After the war's \textbf{election}, \textbf{his} wife \textbf{moved to} \textbf{Dade County, Florida}.                                                                                                                                                                                                                          \\ \cline{2-3}
		             & T5              & \textbf{the} completions \textbf{election}, the was \textbf{his} administration \textbf{moved to} the,\textbf{Dade County, Florida}.                                                                                                                                                                                           \\ \cline{2-3}
		             & ProphetNet      & he the death election \textbf{election}, \textbf{bush} was bush wife \textbf{moved to} washington. \textbf{dade county, florida}.                                                                                                                                                                                              \\ \hline
		
	\end{tabular}
\end{table*}

\section{Discussion}

The results of the BART model and BART + GPT-4 model in both Table~\ref{tab:Table_of_our_results} and Table~\ref{tab:Table_of_our_results} shows that our results is very close to results of paper by Hamza Amrani \cite{Amrani} as we use the same architure, however we have a limited resources of GPU and RAM. The difference that we use other LLMs models which are T5 and ProphetNet to generate sequence but BART prove that is current state-of-the-art model for this task.

Let’s discuss some important definitions:
Precision: Indicates how much of the predicted sentence is relevant to the reference text. 

\begin{equation}
	\text{Precision} = \frac{\text{Number\ of\ correct\ generated\ words}}{\text{Number\ of\ total\ generated\ words}}
	\label{eq:precision}
\end{equation}
Recall: Indicates how much of the key information from the target sentence captured in the predicted sentence.

\begin{equation}
	\text{Re} = \frac{\text{Number\ of\ correct\ generated\ words}}{\text{All\ target\ words}}
	\label{eq:precision}
\end{equation}

F1-score: Uses the harmonic mean of precision and recall providing a balanced metric. Maximizing the F1 score means maximizing both precision and recall.

\begin{equation}
	\text{Recall} = \frac{\text{1}}{\frac{\text{1}}{\text{Precision}} + \frac{\text{1}}{\text{Recall}}}
	\label{eq:precision}
\end{equation}

\textbf{Why we use three evaluation metrics?}

BLEU calculates the number of n-grams (sequences of words) in the predicted sentence match the target sentence exactly. BLEU prioritizes precision: It penalizes the target sentence for adding extra words or slight difference (like "Austin is where he graduated" vs.. "in Austin"), even if the meaning is correct, reducing the BLEU score. Precision indicates how much of the predicted sentence is relevant to the target sentence but using Precision like this is not suitable to use it in this way because it cannot handle repetition and could result in incorrectly perfect precision, as seen in this example Target Sentence: He eats an apple and Predicted Sentence: He He He so the precision = 3/3 = 1. A modified Precision will be used which is Clipped Precision, means if the word in the predicted sentence appears more than 1 time and appears in the target sentence 1 times so this bleu metric will consider as 1 time.

ROUGE-N calculates the number of important phrases and content from target sentence is captured in the predicted sentence, focusing more on recall than precision to know if predicted sentence captures most key ideas from the target (e.g., the university, Austin, Phi Beta Kappa, and the concept of graduating) or not. Even if the phrasing differs (e.g., "at Austin is where" vs. "in Austin"), It recognizes the semantic similarity. It provide precision, recall, and F1-score, measured on a scale from 0 to 1. 

BERTScore measures semantic similarity between the predicted and target sentences, this means it checks if the generated text means the same thing as the original, including both precision (correctness) and recall (completeness). Its precision metric measures how much of the predicted content is semantically aligned with the target.

We explained before in the result section that figure~\ref{fig:bleu_score}, ~\ref{fig:rouge_score}, ~\ref{fig:bert_score} show which model has the highest or lowest BLEU-N score, ROUGE-1 score and BERTScore, In this section we will explain why.

BART + GPT-4 model has best BERTScore as GPT-4 is trained with massive datasets and fine-tuned for high-quality text generation, enabling it to produce more fluent, human-like outputs. GPT-4 can rephrase, restructure, and expand BART’s outputs at a sentence level, reducing redundancy and correcting grammatical errors.

BART(Bidirectional and Auto-Regressive Transformer) model has best  BLEU and ROUGE scores because BART  combines the strengths of both encoder-based models (like BERT) and decoder-based models (like GPT), making it highly effective for tasks such as text generation, summarization, translation, and text classification. In order to detect bi-directional (left and right) dependence, the encoder processes the entire input in parallel, also the model is able to understand word relationships. From left to right, the decoder predicts tokens in a sequential manner, depends on the input sequence and the output that has already been produced, each phase forecasts the next token.

T5 model has a lowest BERTScore because T5 often replaces words with simpler ones or leaves out details as T5 is trained by  removing Chunks of text (instead of just single words) then predicts missing parts, often rewriting them. This makes T5 good at summarizing and for meaning-based tasks, but it may miss details and not capture fine-grained sequence dependencies, so it is not suitable for exact matches. T5 output not similar to the original meaning.

Because ProphetNet prioritizes long-term coherence above word-by-word correctness, it has the lowest BLEU score. A "future n-gram prediction" mechanism is used. This means that during training, it wants to predict several future words (n-grams) at once. Instead of directly replicating words from the input, it paraphrases text, which lowers word overlap and the BLEU score.

The good thing we realize that the integration of pre-trained language models (BART, GPT-4) with EEG signals opens new possibilities for brain-to-text interfaces. This provides a systematic approach and architecture established that can be used as a template or foundation for future work. But unfortunately there are the technical limitations such as the system needs accurate temporal alignment and high-quality EEG data and requires external enhancement using (GPT-4) to optimize output, the training method makes use of sophisticated neural architectures and costly computation resources. Also, there are dataset limitations such as the limited  small number of subjects (30 total across both datasets) and that  subjects were reading in English, limiting generalizability to other languages. Finally there are methodological limitations such as  the current approach requires extensive pre-processing of EEG signals.

\section{Conclusion}

This paper aims to decode the neural activity of high-dimensional data, brain signals recorded by non-invasive electroencephalography (EEG) in the English language. By making sure that the output phrases are coherent and linguistically correct, to improve the quality of the decoded text. Two primary modules were introduced: a brain module and a language modelling module. The Brain module uses a representation learning methodology for EEG encoding. The language modelling module uses BART to generate EEG-to-Text sentences and GPT-4 to refine the sentence level. In training, the first stage emphasises learning EEG features, where the brain module was initially trained to correlate word-level EEG signals with word tokens using Mean Square Error (MSE) loss. To decrease the text reconstruction loss, a pre-trained BART model receives fine-tuning using cross-entropy loss to produce word sequences derived from brain signal representations in the second stage of training.

In the brain module: a) an EEG feature block component, which is a bidirectional Gated Recurrent Unit (GRU), was used to process the time series data to extract the features; b) a subject layer component; to learn a subject-specific vector, it was used to capture inter-subject variability. c) a transformer encoder component, which is a multi-layer transformer encoder, to capture higher-order features; it was used to process the EEG representations and to generate the final latent brain representation. A multi-layer perception (MLP) was followed by processing the EEG representations. In the language modelling module: a) using a pre-trained GPT-4 model via APIs, the text generated by the BART model was refined during inference; b) while preserving the original meaning with minimal changes, the GPT-4 model was promoted to reconstruct the text.

This method enhances the decoding performance as it contributes to: a) a comprehensive deep learning framework for reading open vocabulary EEG signals, b) a representation learning module for raw EEG encoding that is subject-dependent, c) combining a GPT-4 sentence refining module with a BART language model, d) thorough evaluation statistic at the sentence level that is based on the BERTScore, e) an ablation study that examines each module's contributions f) the model's usage of raw EEG signals illustrates the advantages of contemporary representation learning methods in neural science.

Our proposed architecture follows the structure of state-of-the-art models, achieving evaluation metrics: BLEU-1 (42.34\%), BLEU-2 (25.26\%), BLEU-3 (14.91\%), BLEU-4 (8.89\%), ROUGE-1-F (32.66\%), ROUGE-2-F (9.60\%), and BERTScore-F (53.53\%). We trained the architecture using large language models: T5 and ProphetNet. Decoding examples highlight the model's ability to decode named entities, such as "Beta Kappa" and academic achievements, but also reveal limitations, including misinterpretation of historical and political references. Comparison confirms BART's dominance in most metrics and ProphetNet's strength in recall.
    
The results indicate that the BART and BART + GPT-4 models exhibit competitive performance, closely correlating with the findings of Hamza Amrani et al., despite constrained computational resources. BART is the preeminent model for this task, surpassing T5 and ProphetNet in the majority of measures. T5 encounters difficulties with semantic alignment, frequently oversimplifying or neglecting details, whereas ProphetNet's emphasis on long-term coherence diminishes its word-level precision, leading to reduced BLEU scores. BART combined with GPT-4 attains the highest BERTScore owing to GPT-4's capacity to enhance outputs for fluency and semantic accuracy, while BART demonstrates superiority in BLEU and ROUGE scores due to its bidirectional and auto-regressive framework.
    
The combination of pre-trained language models (BART, GPT-4) with EEG inputs offers a potential basis for brain-to-text interfaces. However, obstacles persist, such as the necessity for accurate temporal synchronisation, superior quality EEG data, and significant processing resources. Dataset limitations, including a limited subject quantity and language-specific constraints, limit generality. Although these constraints, the presented framework provides a methodical strategy for upcoming studies in neural decoding and brain-computer interfaces.

The brain signals (EEG) collected during silent reading are decoded into comprehensible English phrases using a deep learning framework presented in this study. The system generates understandable and fluent writing by integrating a two-stage language modelling technique using BART and GPT-4 with a subject-dependent representation learning module for EEG encoding. For people with motor disabilities, this new method has the potential to completely transform communication by allowing more natural and customised human-computer connections for them.

\section{Future Work}

Refining the deep learning architecture can help improve the decoding performance. Decoding handwritten or touch-typing complex movements, according to theoretical considerations, can be easier than point-to-point movements as it enables faster rates of communication. In \cite{Willett2021}, an intracortical brain-computer interface that decodes intended handwriting movements from neural activity in the motor cortex and converts it to text in real-time, giving a recurrent neural network decoding method.

High-performance speech brain-computer interface (BCI) control is known to be made possible by decoding neural activity from the ventral (speech) motor cortex. Previously, it was unclear if this part of the brain, which is usually linked to the dorsal (arm and hand) motor cortex, might also provide computer control through neural cursor and click. The findings of \cite{SingerClark2024} indicate that positioning electrodes in the ventral precentral gyrus (vPCG) to enhance speech decoding could also serve as an effective approach for developing a multi-modal BCI that facilitates both speech communication and computer control through cursor movement and clicking.

By contrasting a machine-translated output with reference translations made by humans, METEOR, an automated machine translation evaluation metric, determines how good the translation is \cite{Banerjee2005}. Its main goal is to match unigrams (single words) between the reference and the machine translation while taking into account a number of matching factors, such as meanings, stems, and surface shapes \cite{Banerjee2005}. Different matching strategies can be accommodated by the framework \cite{Banerjee2005}. Once all potential unigram matches have been found, METEOR combines three factors to determine a score: a) unigram precision (the number of machine translation unigrams that match the reference), b) unigram recall (the number of reference unigrams that the machine translation matches), and c) fragmentation (a measure of how well-ordered the matched words are in the machine translation compared to the reference) \cite{Banerjee2005}. METEOR's efficiency is assessed by looking at how well it correlates with human evaluations of translation quality. The METEOR score has the strongest correlation with human evaluations, and we recommend employing METEOR for task-oriented discussion natural language generation rather than BLEU \cite{Banerjee2005}.

Future research has a great deal of promise to advance electroencephalography (EEG)-based models for brain signal decoding, especially when it comes to speech and cognitive tasks. Compared to modalities like magnetoencephalography (MEG), EEG is more accessible and widely available, although its signals are noisier and less spatially precise \cite{Jayalath2024}. Nevertheless, developments in self-supervised learning (SSL) present encouraging paths around these restrictions \cite{Jayalath2024}. The lack of labelled datasets could be addressed by adopting SSL techniques, which use unlabelled data to pre-train models, to EEG. It would be possible to pre-train models on extensive, unlabelled EEG data in order to teach them strong representations by creating domain-specific pretext tasks, such as those based on neuroscience principles.

In order to allow the model to generalise across participants, tasks, and datasets, these tasks may involve transforming EEG signals to produce implicit training labels. Furthermore, creating data-efficient neural architectures specifically suited to continuous, multi-sensor EEG signals may improve the efficiency of subsequent tasks like classification or voice detection. To find out how increasing the amount of unlabelled data affects model performance, future studies should also investigate scaling laws for EEG-based models, which are comparable to those shown in MEG. EEG-based models may become more accurate and generalisable by combining techniques to integrate data from various studies and participants, opening the door for advances in neurotechnology and brain-computer interfaces.
    
The results of this study have important real-world implications, especially for the creation of assistive technology for people with disabilities. The suggested framework provides a great tool for people with motor limitations, such as: locked-in syndrome, ALS; by decoding EEG signals during silent reading into text while remaining linguistically accurate and coherent. Therefore, people can communicate naturally and intuitively by using text that is produced directly from their brain activity to describe even complicated ideas and feelings.
    
The framework is individualised and efficient due to the subject-dependent representation learning module, which is adaptable to individual brain variability. The decoded text is related linguistically by incorporating sophisticated NLP algorithms, such as BART for sentence formation and GPT-4 for refining. Aside from that, this capability may find usage in virtual reality, gaming, and professional contexts where individuals could operate gadgets or converse solely with their thoughts.

Improvements in BLEU, ROUGE, and BERTScore metrics show the framework's higher performance, which shows the potential for practical implementation. Drawbacks of closed-vocabulary EEG-to-text decoding should be tackled to allow our study for adaptable and versatile brain-to-text systems that can manage open-vocabulary situations and generate good contextually meaningful text. Not only improving communication for people with disabilities but also this development broadens the scope of human-computer interaction and makes using technology easier and more intuitive.

\ifCLASSOPTIONcaptionsoff
\newpage
\fi

\appendices

\section{Website}
\begin{figure*}[!t]
	\renewcommand{\arraystretch}{1.3}
	\centering
	\caption{Signin and Signup pages}
	\label{fig:Signin_and_Signup_pages}
	\includegraphics[width=\textwidth, height=0.4\textheight]{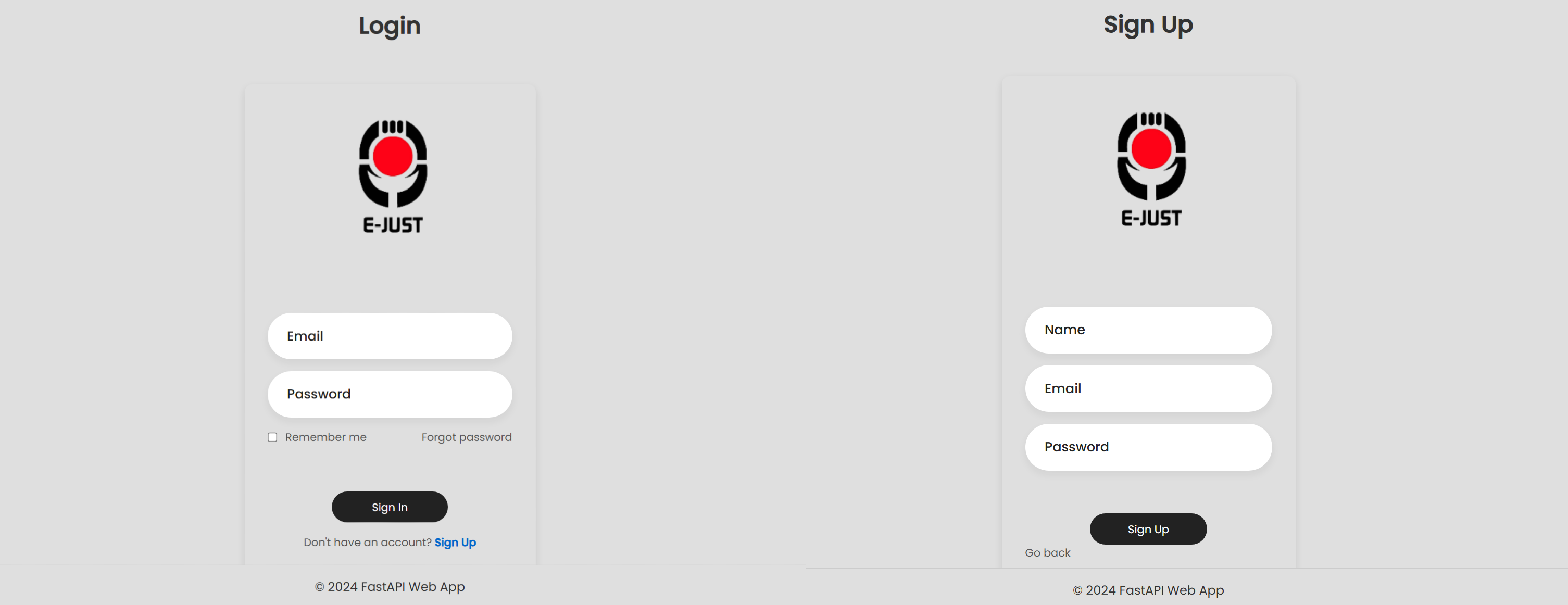} 
	\vspace{-10pt} 
\end{figure*}

A full description of the web-based system is provided in this appendix. The system facilitates both data upload of pickle files and BART and GPT-4 output evaluation to invite user interaction with the decoding system. Figure~\ref{fig:Signin_and_Signup_pages} shows that the users must authenticate their access through registration alongside login features before the platform generates text-to-speech decoded text.

\begin{figure*}[!t]
	\renewcommand{\arraystretch}{1.3}
	\centering
	\caption{Submit page that used to upload the pickel file}
	\label{fig:submit_page}
	\includegraphics[width=0.8\textwidth, height=0.5\textheight]{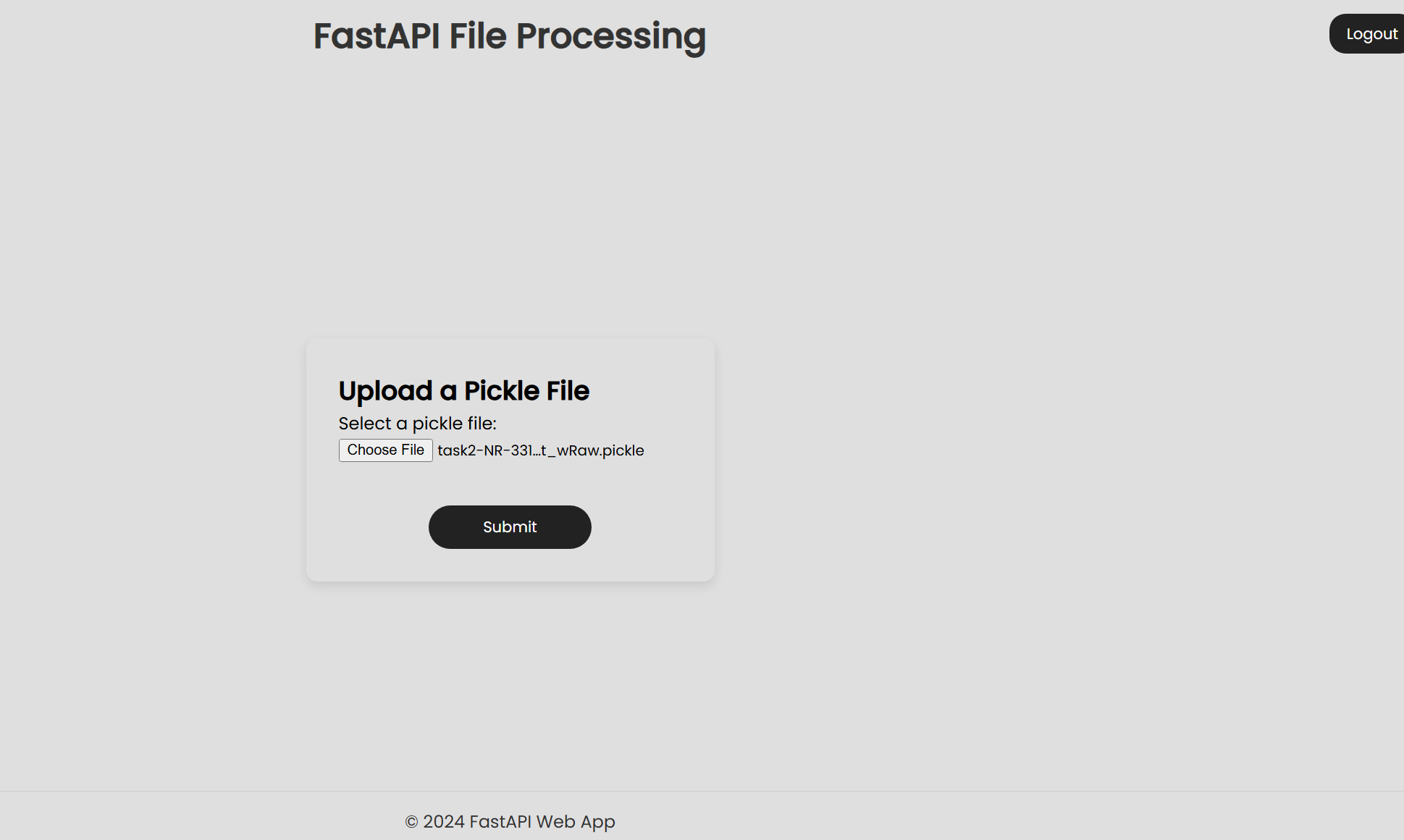} 
	\vspace{-10pt} 
\end{figure*}

\begin{figure*}[!t]
	\renewcommand{\arraystretch}{1.3}
	\centering
	\caption{Predicted and Refined string page and contain the audio files}
	\label{fig:audio_page}
	\includegraphics[width=\textwidth, height=0.5\textheight]{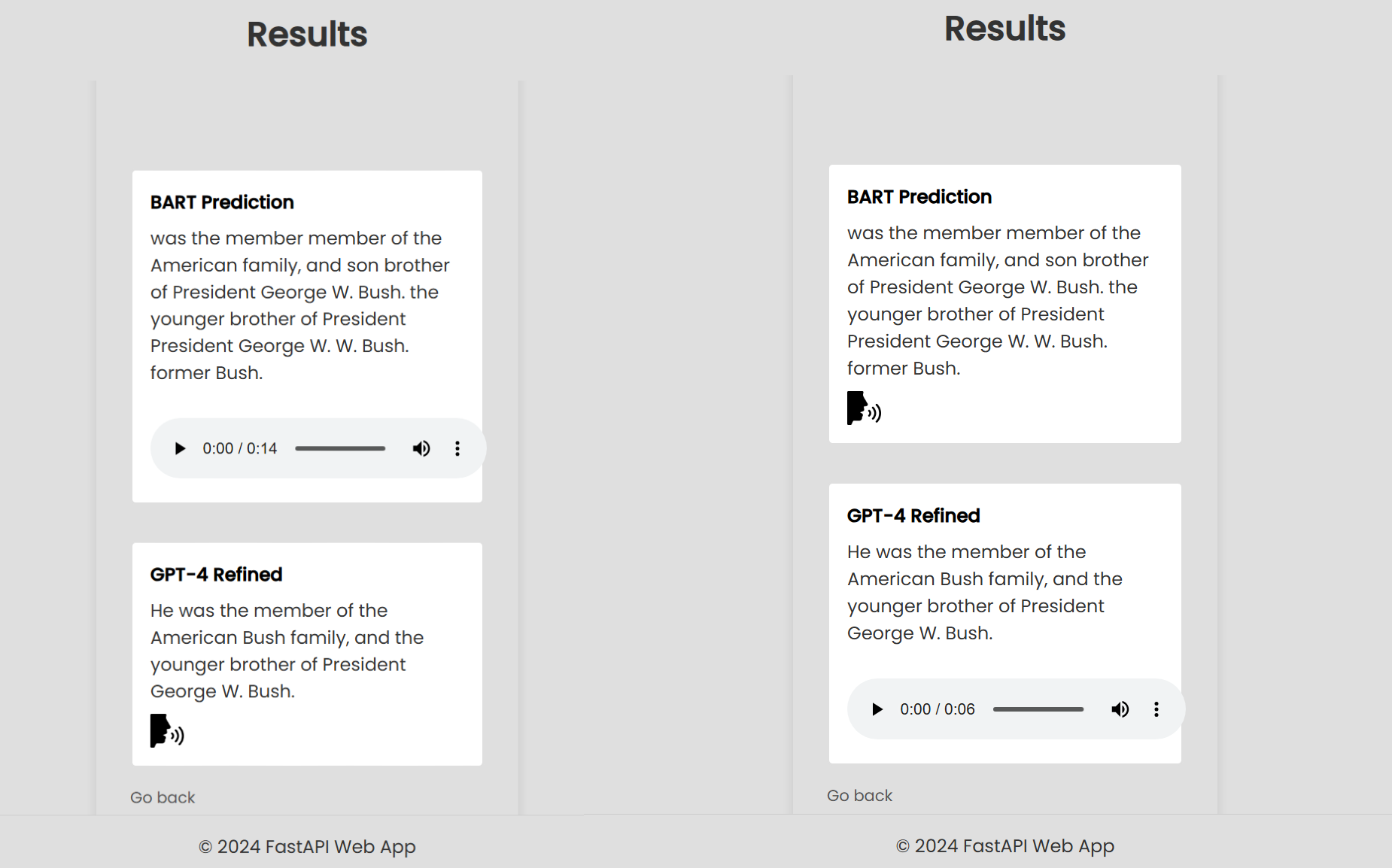} 
	\vspace{-10pt} 
\end{figure*}
\subsection{Project Overview}
The main goal of the EEG to Text Conversion project involves transforming brainwave data (EEG) into readable text through advanced machine learning models. The system allows researchers and users to access its web platform by providing EEG data, which generates readable text. Users experience ease of use during this process.

\subsection{Platform Features}
\begin{enumerate}
	\item To create an account, a user must enter their name and email and set their desired password during registration. The system checks for unique email addresses before permitting new account creation to prevent duplicate accounts.
	\item Registered users can log into the platform using their email address and password. The data upload section becomes available to users after they successfully log into the system.
	\item A pickle file containing EEG data is available for upload by platform users when submitting their data to the system. JMECT applies the GPT-4 and BART models to data processed at this stage.
	\item The results are displayed through two subsections that separate BART Generated Text from GPT-4 Refined Text. BART generates the initial paragraph, while the GPT-4 processing algorithm refines a new text output.
	\item A text-to-speech function integrated within the UI enables users to hear the provided texts. Through its integration of the \texttt{gTTS} (Google Text-to-Speech) software, the platform converts text outputs into spoken audio. Users can directly access the audio through the website after its generation.
	\item A logout option appears within the system, allowing users to return to the home page of the platform following session termination.
\end{enumerate}

\subsection{Frontend Structure}
Users create their interface elements through HTML templates and generate dynamic outputs using the Jinja2 engine. Important frontend elements consist of:

\begin{enumerate}
	\item \texttt{base.html}: Functions as the fundamental design template, generating both header and footer elements while providing CSS styling and the primary content section for the entire website.
	\item \texttt{signup.html}: Allows users to become members by requesting a name, password, and email while displaying field occupation warnings for duplicate email entries.
	\item \texttt{signin.html}: Implements user authentication through an email-address and password form while validating the authentication credentials.
	\item \texttt{index.html}: Enables users to upload pickle file data. Users need to use the logout button for exit as shown in Figure~\ref{fig:submit_page}.
	\item \texttt{results.html}: Displays decoded text, text-to-speech controls, and BART and GPT-4 responses.
	\item \texttt{index2.html}: Provides access to the text entry page featuring the text-to-speech functionality as shown in Figure~\ref{fig:audio_page}.
\end{enumerate}

\subsection{Backend Architecture}
The backend implements its structure using FastAPI as its Python framework. The application handles user verification, data input, EEG signal processing, and text-to-speech output generation. Key backend components consist of:

\begin{enumerate}
	\item A SQLite database stores all information regarding user management, including username, email address, and password. The demonstration storage of passwords in plain text requires migration to hashing methods for production security.
	\item During registration, the system checks for duplicate emails while validating user-provided login credentials.
	\item GPT-4 and BART models function within the data management system for pickle file uploading and processing. The generated text appears as an output that gets saved within the system.
	\item Through the \texttt{gTTS} library, the application converts text output into audio files, which get saved in the \texttt{static} directory before being forwarded to the frontend.
	\item The FastAPI session management tracks all users who have active logins. The session period ends at logout time.
	\item The system handles upload errors, authentication issues, and speech synthesis errors by delivering essential updates to users.
\end{enumerate}

\subsection{Deployment Strategy}
The FastAPI application receives its public URL from \textbf{ngrok} when conducting platform launches. Ngrok creates a channel that connects to the server operating on port 8000 in the local system.

\subsection{Summary}
Users can access the decoding system through the intuitive interface of the web application EEG to Text Conversion. The platform offers multiple functionalities, including user authentication, data upload, decoded text, and text-to-speech features. The backend runs FastAPI, while the frontend relies on HTML together with Jinja2 for handling dynamic content. The platform demonstrates the practical functionality of the EEG to Text Conversion project through its operational methods.


\begin{thebibliography}{1}
	
	\bibitem{unknown-author-2022A}
	Unknown Author, ``Aphasia Statistics - The National Aphasia Association,'' Mar. 2022. [Online]. Available: \url{https://aphasia.org/aphasia-resources/aphasia-statistics/}.
	
	\bibitem{lyberg-ahlander-2018}
	V. Lyberg-Åhlander, R. Rydell, P. Fredlund, C. Magnusson, and S. Wilén, ``Prevalence of voice disorders in the general population, based on the Stockholm Public Health cohort,'' \textit{Journal of Voice}, vol. 33, no. 6, pp. 900--905, Aug. 2018. [Online]. Available: \url{https://doi.org/10.1016/j.jvoice.2018.07.007}.
	
	\bibitem{le-h-2025}
	H. Le, F. Lui, and M. Y. Lui, \textit{Aphasia}. StatPearls Publishing, 2025. [Online]. Available: \url{https://www.ncbi.nlm.nih.gov/books/NBK559315/}.
	
	\bibitem{unknown-author-2022B}
	Unknown Author, ``Public Figures and Celebrities with Aphasia - The National Aphasia Association,'' Mar. 2022. [Online]. Available: \url{https://aphasia.org/aphasia-resources/public-figures-and-celebrities-with-aphasia/}.
	
	\bibitem{servick-2021}
	K. S. Servick, ``Brain signals converted into words ‘speak' for person with paralysis,'' Jul. 2021. [Online]. Available: \url{https://www.science.org/content/article/brain-signals-converted-words-speak-person-paralysis}.
	
	\bibitem{Moses2021} 
	D. A. Moses, S. L. Metzger, J. R. Liu, G. K. Anumanchipalli, J. G. Makin, P. F. Sun, J. Chartier, M. E. Dougherty, P. M. Liu, G. M. Abrams et al., ``Neuroprosthesis for decoding speech in a paralyzed person with anarthria,'' \textit{New England Journal of Medicine}, vol. 385, no. 3, pp. 217–227, 2021.
	
	\bibitem{Willett2023} 
	F. R. Willett, E. M. Kunz, C. Fan, D. T. Avansino, G. H. Wilson, E. Y. Choi, F. Kamdar, L. R. Hochberg, S. Druckmann, K. V. Shenoy et al., ``A high-performance speech neuroprosthesis,'' \textit{bioRxiv}, 2023.
	
	\bibitem{Hollenstein2018} 
	N. Hollenstein, J. Rotsztejn, M. Troendle, A. Pedroni, C. Zhang, and N. Langer, ``ZuCo, a simultaneous EEG and eye-tracking resource for natural sentence reading,'' \textit{Scientific Data}, vol. 5, no. 1, pp. 1–13, 2018.
	
	\bibitem{Panachakel2021} 
	J. T. Panachakel and A. G. Ramakrishnan, ``Decoding covert speech from EEG—A comprehensive review,'' \textit{Frontiers in Neuroscience}, vol. 15, p. 392, 2021.
	
	\bibitem{Duan2023} 
	Y. Duan, J. Zhou, Z. Wang, Y.-K. Wang, and C.-T. Lin, ``DeWave: Discrete EEG waves encoding for brain dynamics to text translation,'' \textit{arXiv preprint arXiv:2309.14030}, 2023.
	
	\bibitem{Feng2023} 
	X. Feng, X. Feng, and B. Qin, ``Semantic-aware contrastive learning for electroencephalography-to-text generation with curriculum learning,'' \textit{arXiv preprint arXiv:2301.09237}, 2023.
	
	\bibitem{Wang2022} 
	Z. Wang and H. Ji, ``Open vocabulary electroencephalography-to-text decoding and zero-shot sentiment classification,'' in \textit{Proceedings of the AAAI Conference on Artificial Intelligence}, vol. 36, 2022, pp. 5350–5358.
	
	\bibitem{Pandarinath2017} 
	C. Pandarinath, P. Nuyujukian, C. H. Blabe, B. L. Sorice, J. Saab, F. R. Willett, L. R. Hochberg, K. V. Shenoy, and J. M. Henderson, ``High performance communication by people with paralysis using an intracortical brain-computer interface,'' \textit{eLife}, vol. 6, p. e18554, 2017.
	
	\bibitem{Jeng2020} 
	P.-Y. Jeng, C.-S. Wei, T.-P. Jung, and L.-C. Wang, ``Low-dimensional subject representation-based transfer learning in EEG decoding,'' \textit{IEEE Journal of Biomedical and Health Informatics}, vol. 25, no. 6, pp. 1915–1925, 2020.
	
	\bibitem{Devlin2018} 
	J. Devlin, M.-W. Chang, K. Lee, and K. Toutanova, ``BERT: Pre-training of deep bidirectional transformers for language understanding,'' \textit{arXiv preprint arXiv:1810.04805}, 2018.
	
	\bibitem{Lewis2019} 
	M. Lewis, Y. Liu, N. Goyal, M. Ghazvininejad, A. Mohamed, O. Levy, V. Stoyanov, and L. Zettlemoyer, ``BART: Denoising sequence-to-sequence pre-training for natural language generation, translation, and comprehension,'' \textit{arXiv preprint arXiv:1910.13461}, 2019.
	
	\bibitem{Caucheteux2022} 
	C. Caucheteux and J.-R. King, ``Brains and algorithms partially converge in natural language processing,'' \textit{Communications Biology}, vol. 5, no. 1, p. 134, 2022.
	
	\bibitem{Tang2023} 
	J. Tang, A. LeBel, S. Jain, and A. G. Huth, ``Semantic reconstruction of continuous language from non-invasive brain recordings,'' \textit{Nature Neuroscience}, pp. 1–9, 2023.
	
	\bibitem{OpenAI2023} 
	OpenAI, GPT-4 Technical Report, \textit{arXiv:2303.08774 [cs.CL]}, 2023. Available: \url{https://arxiv.org/abs/2303.08774}
	
	\bibitem{Gauthier2018} 
	J. Gauthier and A. Ivanova, ``Does the brain represent words? An evaluation of brain decoding studies of language understanding,'' \textit{arXiv preprint arXiv:1806.00591}, 2018.
	
	\bibitem{Nieto2022} 
	N. Nieto, V. Peterson, H. L. Rufiner, J. E. Kamienkowski, and R. Spies, ``Thinking out loud, an open-access EEG-based BCI dataset for inner speech recognition,'' \textit{Scientific Data}, vol. 9, no. 1, p. 52, 2022.
	
	\bibitem{Vaswani2017} 
	A. Vaswani, N. Shazeer, N. Parmar, J. Uszkoreit, L. Jones, A. N. Gomez, Ł. Kaiser, and I. Polosukhin, ``Attention is all you need,'' in \textit{Advances in Neural Information Processing Systems}, 2017, pp. 5998–6008.
	
	\bibitem{Lin2004} 
	C.-Y. Lin, ``ROUGE: A package for automatic evaluation of summaries,'' in \textit{Text Summarization Branches Out}, 2004, pp. 74–81.
	
	\bibitem{Zhang2019} 
	T. Zhang, V. Kishore, F. Wu, K. Q. Weinberger, and Y. Artzi, ``BERTScore: Evaluating text generation with BERT,'' \textit{arXiv preprint arXiv:1904.09675}, 2019.
	
	\bibitem{Cho2014} 
	K. Cho, B. Van Merrienboer, D. Bahdanau, and Y. Bengio, ``On the properties of neural machine translation: Encoder-decoder approaches,'' \textit{arXiv preprint arXiv:1409.1259}, 2014.
	
	\bibitem{Jarosiewicz2015} 
	B. Jarosiewicz, A. A. Sarma, D. Bacher, N. Y. Masse, J. D. Simeral, B. Sorice, E. M. Oakley, C. Blabe, C. Pandarinath, V. Gilja et al., ``Virtual typing by people with tetraplegia using a self-calibrating intracortical brain-computer interface,'' \textit{Science Translational Medicine}, vol. 7, no. 313, pp. 313ra179–313ra179, 2015.
	
	\bibitem{Willett2021} 
	F. R. Willett, D. T. Avansino, L. R. Hochberg, J. M. Henderson, and K. V. Shenoy, ``High-performance brain-to-text communication via handwriting,'' \textit{Nature}, vol. 593, no. 7858, pp. 249–254, 2021.
	
	\bibitem{Lee2018} 
	M.-H. Lee, J. Williamson, D.-O. Won, S. Fazli, and S.-W. Lee, ``A high performance spelling system based on EEG-EOG signals with visual feedback,'' \textit{IEEE Transactions on Neural Systems and Rehabilitation Engineering}, vol. 26, no. 7, pp. 1443–1459, 2018.
	
	\bibitem{Makin2020} 
	J. G. Makin, D. A. Moses, and E. F. Chang, ``Machine translation of cortical activity to text with an encoder-decoder framework,'' \textit{Nature Neuroscience}, vol. 23, no. 4, pp. 575–582, 2020.
	
	\bibitem{Anumanchipalli2019} 
	G. K. Anumanchipalli, J. Chartier, and E. F. Chang, ``Speech synthesis from neural decoding of spoken sentences,'' \textit{Nature}, vol. 568, no. 7753, pp. 493–498, 2029.
	
	\bibitem{Brigham2010} 
	K. Brigham and B. V. Kumar, ``Imagined speech classification with EEG signals for silent communication: A preliminary investigation into synthetic telepathy,'' in \textit{2010 4th International Conference on Bioinformatics and Biomedical Engineering}, 2010, pp. 1–4.
	
	\bibitem{Defossez2023} 
	A. Defossez, C. Caucheteux, J. Rapin, O. Kabeli, and J.-R. King, ``Decoding speech perception from non-invasive brain recordings,'' \textit{Nature Machine Intelligence}, pp. 1–11, 2023.
	
	\bibitem{Amrani} 
	H. Amrani, D. Micucci, and P. Napoletano, ``Deep Representation Learning for Open Vocabulary Electroencephalography-to-Text Decoding,'' \emph{IEEE Journal of Biomedical and Health Informatics}, 2024.
	
	\bibitem{SingerClark2024} 
	T. Singer-Clark \textit{et al.}, “Speech motor cortex enables BCI cursor control and click,” 
	\textit{bioRxiv (Cold Spring Harbor Laboratory)}, Nov. 2024, doi: 10.1101/2024.11.12.623096.
	
	\bibitem{Banerjee2005} 
	S. Banerjee and A. Lavie, “METEOR: An Automatic Metric for MT Evaluation with Improved Correlation with Human Judgments,” 
	\textit{ACL Anthology}, Jun. 01, 2005. \url{https://aclanthology.org/W05-0909/}.
	
	\bibitem{Jayalath2024} 
	D. Jayalath, G. Landau, B. Shillingford, M. Woolrich, and O. P. Jones, “The brain’s bitter lesson: 
	Scaling speech decoding with Self-Supervised Learning,” \textit{arXiv.org}, Jun. 06, 2024. 
	\url{https://arxiv.org/abs/2406.04328}.
	
\end{thebibliography}
\end{document}